\documentclass[journal]{IEEEtran}
\usepackage[ruled,vlined]{algorithm2e}
\usepackage{color}
\usepackage{epsf}
\usepackage{times}
\usepackage{epsfig}
\usepackage{graphicx}
\usepackage{url}
\usepackage{amsmath}
\usepackage{amssymb}
\usepackage{amsxtra}
\usepackage{cite}
\usepackage{listings}
\usepackage{verbatim}
\usepackage{pseudocode}
\usepackage{mathrsfs}
\usepackage{float}
\usepackage{subfig}
\usepackage{epstopdf}

\newtheorem{assumption}{Assumption}

\newtheorem{theorem}{\bf Theorem}[section]

\newtheorem{lemma}{Lemma}[section]
\newtheorem{example}{Example}[section]

\newtheorem{remark}{Remark}
\newtheorem{problem}{Problem}[section]
\begin{document}
\title{Randomized Consensus based Distributed Kalman Filtering over Wireless Sensor Networks}

\author{Jiahu~Qin,~\IEEEmembership{Senior~Member,~IEEE,}
	Jie~Wang,~\IEEEmembership{Student~Member,~IEEE,}
	Ling~Shi,~\IEEEmembership{Senior~Member,~IEEE,}
	and~Yu~Kang,~\IEEEmembership{Senior~Member,~IEEE}
	\thanks{J. Qin and J. Wang are with the Department of Automation, University
		of Science and Technology of China, Hefei 230027, China (e-mail:
		jhqin@ustc.edu.cn; wj1993@mail.ustc.edu.cn).}
	\thanks{L. Shi is with the Department of Electronic and Computer Engineering,
		Hong Kong University of Science and Technology, Clear Water Bay, Kowloon,
		Hong Kong, China (e-mail: eesling@ust.hk).}
	\thanks{Y. Kang is with the Department of Automation, State Key Laboratory
		of Fire Science, and the Institute of Advanced Technology, University of
		Science and Technology of China, Hefei 230027, China, and also with
		the Key Laboratory of Technology in Geo-Spatial Information Processing
		and Application System, Chinese Academy of Sciences, Beijing 100190,
		China (e-mail: kangduyu@ustc.edu.cn).}
}

\maketitle

\begin{abstract}
   	This paper is concerned with developing a novel distributed Kalman filtering algorithm over wireless sensor networks based on randomized consensus strategy. 
    Compared with centralized algorithm, distributed filtering techniques require less computation per sensor and lead to more robust estimation since they simply use the information from the neighboring nodes in the network.
    However, poor local sensor estimation caused by limited observability and network topology changes which interfere the global consensus are challenging issues.  Motivated by this observation, we propose a novel randomized gossip based distributed Kalman filtering algorithm. Information exchange and computation in the proposed algorithm can be carried out in an arbitrarily connected network of nodes. In addition, the computational burden can be distributed for a sensor which communicates with a stochastically selected neighbor at each clock step under schemes of gossip algorithm.
In this case, the error covariance matrix changes stochastically at every clock step, thus the convergence is considered in a probabilistic sense. We provide the mean square convergence analysis of the proposed algorithm. Under a sufficient condition, we show that the proposed algorithm is quite appealing as it achieves better mean square error performance theoretically than the noncooperative decentralized Kalman filtering algorithm. Besides, considering the limited computation, communication and energy resources in the wireless sensor networks, we propose an optimization problem which minimizes the average expected state estimation error based on the proposed algorithm. To solve the proposed problem efficiently, we transform it into a convex optimization problem. And a sub-optimal solution is attained. Examples and simulations are provided  to illustrate the theoretical results.
\end{abstract}

\section{Introduction}\label{sec:intro}
A wireless sensor network (WSN) is a network composed of a large number of sensor nodes where each node is equipped
with processing, communication and sensing capabilities.
All the nodes work cooperatively to monitor physical or environmental conditions, such as temperature, sound and pressure. The development of WSNs are motivated by military and environmental applications, such as battlefield surveillance and flood detection\cite{Akyildiz2002}. Today such networks are used in widescope applications such as robotics, surveillance, smart grid and health care, etc \cite{Fax2004,Kim2012}.

Multi-sensor estimation through WSN gives higher accuracy than a single sensor node \cite{Saber2007}. Thus, a completely centralized Kalman filtering algorithm is proposed in which all the observations of the sensors are sent to a
central processing facility to perform the global data fusion.
However, a fine grained measurement setup will result in a large amount of information that requires further processing
and communication\cite{Bertsekas1999}.  Due to the limited energy, communication, computation and storage resources, it may be impossible for all the sensors to send their observations to a central unit. Decentralized Kalman filtering iteration \cite{Yang2013} involves state estimation using the measurement of local neighboring nodes
in which there is no centralized processing station.
Different from the local decentralized Kalman filtering algorithm, the distributed Kalman filtering iteration incorporates the priori estimates of the neighbors into the estimation.  Incorporation of neighboring states brings the opportunity to reach consensus throughout the network inherently on the state estimation.
The distributed Kalman filtering algorithm is proposed in which each node estimates the state by communicating with its neighbors and then reaches consensus on the system state.  Under distributed Kalman filtering, a sensor does not have to transmit data to a center and can share the information with its neighbors to increase the reliability of the estimation and ensure consistency.


Consensus algorithms are powerful tools to carry out network-wide distributed computation tasks such as computing aggregate quantities and functions over networks \cite{Saber2007,Saber2004,Kempe2003,Boyd2005}.
The work of \cite{Saber2005a} focused on dynamic distributed sensor fusion to obtain consensus
weighted least-squares fused estimates for multiple measurements. However, \cite{Saber2005a} does not include the
system dynamic equations of targets and no direct connection is obtained with the Kalman filter. The work in \cite{Saber2005} proposed some distributed Kalman filtering algorithms, which include a network of micro-Kalman filters that embedded with a low-pass and a band-pass consensus filter. The Kalman consensus filtering algorithm first proposed in \cite{Saber2007(2)} has proved to be
a popular and influential distributed consensus-based framework
for dynamic state estimation. A formal stability and performance analysis of Kalman consensus filtering algorithm was given
in \cite{Saber2009}. The consensus terms in these papers
are added in an ad hoc fashion outside the Kalman filter
framework. Cattivelli and Sayed \cite{Cattivelli2010} proposed diffusion strategies for distributed filtering and smoothing. Simulation shows improved performance relative to the Kalman consensus filtering algorithm \cite{Saber2009}. Yu \textit{et al.} \cite{Yu2009} proposed a distributed consensus filtering algorithm based on pinning control, where only a small fraction of sensors need to measure the target information, with which the whole network can be controlled.

One of the key consideration in WSNs is that both
computation and transmission are time and energy consuming
tasks which have to be reduced as much as possible.
 Strategies for distributed estimation \cite{Cattivelli2010,Cui2013,Demetriou2010,Spanos2005} are not closely related to the network topology and can reduce the rate of communication to cut down the energy cost. Besides, it is more flexible for ad-hoc deployment when compared with centralized and decentralized estimations\cite{Rao1993,Sanders1974}. However, there are also some challenges when adopting the distributed estimation strategy in WSNs. For example, the energy of sensors distributed in a complex environment is usually limited and the battery power is difficult to be regained or supplied. Thus, for a sensor, effectively selecting some of its neighboring sensors to send data to can extend the lifetime of its power source while guaranteeing a desired level of estimation quality.

Many literatures \cite{Saber2007(2),Cattivelli2010,Kamal2011,Kamal2013,Wang2014,Ji2017} have considered only the case of  a stationary communication strategy which is described by
a constant consensus matrix. However, this is a simple model of communications in many practical applications.
For example, the WSNs need to frequently adjust the network topology due to dynamical operating environments.
In this case the adopted randomized gossip strategy can provide robustness with respect to dynamic environments,
even if network topology is subject to frequent and unpredictable variations.
The resilience to network topology changes makes gossip
protocols particularly appealing as an algorithmic framework for the distributed averaging problem in a dynamic setting.  However, their rate of convergence is an issue. In fact,  large number of iterations required to achieve target accuracy affect the energy budget and, in turn, the network lifetime. Therefore, several efforts have been made to improve convergence speed and mitigate energy consumption\cite{Freschi2016},\cite{Dimakis2006,Dimakis2008}.
 Besides, the work in \cite{Li2017} investigates sensor transmission power control for remote state estimation. Instead of using a conventional sensor, a sensor equipped with an energy harvester which can obtain energy from the external environment is utilized.
Also it should be pointed out that in many practical applications a node cannot simultaneously receive data from two different
neighbor nodes (for example collision can destroy messages in wireless environment) and in some applications data cannot
simultaneously be transmitted to more than one node \cite{Fagnani2008}.  This
fact makes the use of randomized consensus algorithms quite appealing as it picks up randomly one of its neighboring nodes and exchanges its estimate.
It turns out that randomized consensus strategies can achieve better performance than deterministic ones with comparable complexity.

In this paper, we propose the randomized gossip based distributed Kalman filtering algorithm which is robust against changes in topology. This method computes a local state estimate using the measurement data from the neighborhoods of every node. Subsequently, in the randomized consensus step one node randomly wakes up, picks up randomly one of its neighbor nodes and exchanges its estimate.
Our main technical contribution is to provide a formal stability and performance analysis of the proposed algorithm and show that the estimation performance is better when compared with $n$ noncooperative decentralized Kalman filtering algorithm. The use of randomized protocols avoids the need of cumbersome communication scheduling, reduces the need of time synchronization and may also reduce power consumption. A further cause of randomness in the communication is the potential unpredictability of the environment where these protocols are implemented: packet losses and collisions are in fact rather common in a sensor network. Finally, we consider optimal sensor scheduling for distributed estimation subject to limited power.
The main contributions of this paper are summarized as follows:
\begin{itemize}
\item [1)] Motivated by the results obtained in  \cite{Saber2007(2)},  we derive the distributed implementation of the centralized Kalman filtering algorithm. The difference compared with \cite{Saber2007(2)} is that we adopt the randomized consensus algorithm for fusion of sensor data and covariance information.
\item [2)] Different from the works in \cite{Cattivelli2010} and \cite{Saber2007(2)} which focus on average consensus, we propose a novel distributed Kalman filtering algorithm based on randomized consensus algorithm which is robust against the changes of the network topology. Convergence is considered in the probabilistic sense. We provide a rigorous stability analysis which is one of the main technical contributions of this work. Under a sufficient condition that $\mathcal{P}(\mathcal{P}^{-})^{-1}(I \otimes A)$ is an orthogonal matrix, we show that the proposed algorithm is quite appealing as it has better performance than noncooperative decentralized Kalman filtering algorithm.
\item [3)]  Considering the limited communication resource in  WSNs, we provide a sub-optimal sensor scheduling scheme for distributed estimation subject to limited power.
\end{itemize}

The remainder of the paper is organized as follows. In section \ref{sec:problem setup}, we first introduce the system model, graph theory preliminary and randomized consensus algorithm. The distributed implementation of the centralized estimation based on the randomized consensus algorithm is then developed. In section~\ref{sec:RGDF}, we formulate the randomized gossip based distributed Kalman filtering algorithm and provide a formal stability and performance analysis. The optimal sensor scheduling for our distributed estimation subject to limited power is formulated in section \ref{sec:Problem statement}.
Section \ref{sec:simulation} presents a numerical example to illustrate the performance of the optimal sensor scheduling scheme and provides the performance comparison of different distributed Kalman filterings algorithms to show the optimality of our algorithm.


\textbf{Notations:}
$Z$ is the set of non-negative integers. $k$ is the time index. $N$ is the set of natural numbers. $R^{n}$ in $n$-dimensional Euclidian
spaces. $S^{n}_{+}$ and $S^{n}_{++}$ are the sets of $n\times n$
 positive semi-definite and positive-definite matrices, respectively. When $X\in S^{n}_{+}$, we simply write $X\geq 0$ or $X>0$ if $X\in S^{n}_{++}$.
 For a matrix $X$, $X'$ denotes its transpose. $Tr[\cdot]$ denotes the trace of a matrix. $X\geq Y$ if $X-Y \in S^{n}_{+}$. $E[\cdot]$ denotes the expectation of a random
 variable. For function $f_{1},f_{2}$ with appropriate domains, $f_{1}f_{2}(x)$ denotes the function composition $f_{1}(f_{2}(x))$,
 and $f^{n}(x)\triangleq f(f^{n-1}(x))$ with $ f^{0}\triangleq x$.
 Symbol $\otimes$ represents the Kronecker product, $\lambda_{2}(M)$ denotes the second largest eigenvalues of matrix $M$.
\section{PRELIMINARIES And Problem Formulation}\label{sec:problem setup}
\subsection{System Model}
    Considering the following discrete linear time-invariant system:
    \begin{align}\label{systemdynamic}
        x(k+1)=Ax(k)+w(k),
    \end{align}
    where $x(k)\in R^{m}$ is the system state vector at time $k$, $w(k)$ is the process noise. Assume that $x(0)$ and $w(k)$ are independent zero-mean Gaussian random vectors with covariances $\Pi_{0}$ and $Q$, respectively.
    A sensor network composed of $n$ sensors is used to measure the system state $x(k)$. The measurement equation of the $i$th sensor is given by
    \begin{align}
        y_{i}(k)=C_{i}x(k)+v_{i}(k), i=1,2,...,n,
    \end{align}
    where $v_{i}(k)\in R^{m_{i}}$ is zero-mean white Gaussian with covariance matrix $R_{i}>0$ which is independent of $x(0), w(k)$, for $\forall k , i,$ and is independent of $v_{j}(s)$ when $i \neq j$ or $k \neq s$. Hence, we have
    \begin{equation}
    \begin{aligned}
    &E\{[w(k), v_{i}[k], v_{j}[k]]^{T}[w(k), v_{i}[k], v_{j}[k]]\}\\
   	&=diag\{Q, R_{i}, R_{j}\}\delta_{kt}, i,j=1,...,n, i \neq j,
   	\end{aligned}
   	\end{equation} where $\delta_{kk}=1$ and $\delta_{kt}=0$ for $ k \neq t.$
    The pair $(A ,C_{i})$ is assumed to be observable and $(A,\sqrt{Q})$ is controllable.
\subsection{Graph Theory Preliminaries}
    We model the sensor network as a undirected graph $G=(\mathcal{V},\varepsilon)$ with the nodes $\mathcal{V}=\{1,2,....n\}$ being the sensors and the edges $\varepsilon \subset \mathcal{V} \times \mathcal{V}$ representing the communication links.
    We define the adjacent matrix $ \Gamma =[\gamma_{ij}] $ as follows. When $\gamma_{ij}=1$, there exists an edge $(i,j)$ representing that the $j$th  node receives data from the $i$th one, and $\gamma_{ij}=0$ indicates that the $j$th  node does not receive data from $i$th one. Since a sensor node is always able to access its own observed values, we have $\gamma_{ii}=1$.
    We consider undirected graph in this paper. The adjacent matrix of an undirected graph is defined as $ \gamma_{ij}=\gamma_{ji}$ if $(j,i)\in \varepsilon,$ where $i\neq j$.
    The set of incoming neighbors to a node $v_{i}$ is given as $\mathcal{N}_{i}=\{j \in \mathcal{V}: \gamma_{ji}=1\}$, and the set of outgoing neighbors is defined as $O_{i}=\{j \in \mathcal{V}: \gamma_{ij}=1\}$. The in-degree of node $i$ denoted as $d^{I}_{i}$ is given by $d^{I}_{i}=|\mathcal{N}_{i}|$. Similarly, the out-degree $d^{O}_{i}$ of node $i$ is given by $d^{O}_{i}=|O_{i}|$.
\subsection{Estimation Algorithm}
    We propose some notations which will be used in the remainder of the paper, where the expectations are taken with respected to the observed values and process noise $v_{i}(k),w(k)$ as follows:
    \begin{equation}
    \zeta^{i}_{k} \triangleq \{y_{i}(0),y_{i}(1),...,y_{i}(k)\}
    \end{equation}
    with $\zeta^{i}_{-1}\triangleq \emptyset$, $i=1,2,...n$. Furthermore, let
    \begin{eqnarray}
    \hat{x}^{i}_{k|k-1} &\triangleq& E[x^{i}_{k}|\zeta^{i}_{k-1}],\\
   P^{i}_{k|k-1} &\triangleq&  E[(x_{i}-\hat{x}^{i}_{k|k-1})(x_{i}-\hat{x}^{i}_{k|k-1})^{'}|\zeta^{i}_{k-1}],\\
    \hat{x}^{i}_{k} &\triangleq& E[x^{i}_{k}|\zeta^{i}_{k}],\\
    P^{i}_{k} &\triangleq&  E[(x^{i}_{k}-\hat{x}^{i}_{k})(x^{i}_{k}-\hat{x}^{i}_{k})^{'}|\zeta^{i}_{k}].
    \end{eqnarray}
    where $P^{i}_{k|k-1}$ and $P^{i}_{k}$ denote the state covariance matrices and their inverses are known as the \textit{information matrices.} Note that $\hat{x}^{i}_{0|-1}=0$ and $P^{i}_{0|-1}=\Pi_{0}$. Here are the Kalman filtering iterations in the information form:
    \begin{eqnarray}
     (P^{i}_{k})^{-1}&=&(P^{i}_{k|k-1})^{-1}+C_{i}'R^{-1}_{i}C_{i},\\
     K^{i}_{k}&=&P^{i}_{k}C_{i}'R^{-1}_{i}, \\
     \hat{x}^{i}_{k}&=&\hat{x}^{i}_{k|k-1}+ K^{i}_{k}(y_{i}(k)-C_{i}\hat{x}^{i}_{k|k-1}),\\
     P^{i}_{k+1|k}&=&AP^{i}_{k}A'+Q,\\
     \hat{x}^{i}_{k+1|k}&=&A\hat{x}^{i}_{k}.
    \end{eqnarray}
\subsection{Randomized Gossip Algorithm}
In the following, the randomized gossip algorithm which is used for reaching a consensus on the local estimates against changes in topology is introduced.
First, we define a stochastic matrix $P =[P_{ij}]$ of nonnegative entries with the condition that $P_{ij}>0$ only if $(i,j) \in \mathcal{V} \times \mathcal{V}$. For technical reasons, we assume that $P$ is a stochastic matrix.
 We define a set of stochastic matrices as follows:
\begin{align}\label{15}
W_{ij}=I-\frac{(e_{i}-e_{j})(e_{i}-e_{j})^{'}}{2},
\end{align}
where $e_{i}=[0 ... 0 , 1 , 0...0]'$ is an $n \times 1$ unit vector with the $i$th component equal to 1.
Formally, let $\xi(t)$ denote the vector of state values at the end of the time-slot $t$. We have:
$
\xi(t+1)=W(t)\xi(t),
$
where the random matrix $W(t)$, with probability $\frac{1}{n}P_{i,j}$
is randomly selected from the set whose elements are defined in \eqref{15}, that is $W(t)=W_{ij}$.
According to \cite{Boyd2006},  if $\xi(t)$ converges to the vector of averages $\xi_{ave}\textbf{1}=\frac{\boldsymbol{1}\boldsymbol{1}^{T}}{n}\xi(0)$, we must have
$
\lim \limits_{t\rightarrow{\infty}}\Phi(t)=\lim\limits_{t\rightarrow{\infty}}W(t)W(t-1)\cdots W(0)=\boldsymbol{1}\boldsymbol{1}^{T}/n
$
for every initial condition $\xi(0)$.


To obtain the convergence of $\xi(t)$ to $\xi_{\infty}$, we will consider the convergence of consensus error defined as $e(t)=\xi(t)-\xi(\infty)$.
Thus, $e(k)$ evolves according to the same linear system as $\xi$.
Then, repeatedly conditioning and using the linear iteration, the following inequality holds \cite{Boyd2005}:
\begin{equation}\label{18}
E[e(k)^{T}e(k)]\leq \lambda_{2}^{k}(E[W^{T}W])||e(0)||^{2}_{2}.
\end{equation}
From this, we can see that the second moment of the error $e(k)$ converges to $0$ at a rate governed by $\lambda_{2}(E[W^{T}W])$.
This means that any scheme of choosing the $W(t)$ with second largest eigenvalue strictly less than 1 (and, of course, with $\rho(E(W)-\boldsymbol{1}\boldsymbol{1}^{T}/n)$ less than 1 \cite{Boyd2006}) is convergent in the second moment.

In \cite{Boyd2006} the authors also propose the $\varepsilon$-averaging time of the randomized consensus algorithm denoted by
$T_{ave}(\varepsilon)$ which is defined as
\begin{equation}
\sup \limits_{\xi(0)}\inf\Bigg\{t:Pr\Big[\frac{||\xi(t)-\xi_{ave}\textbf{1}||}{||\xi(0)||}\geq \varepsilon\Big]\leq \varepsilon\Bigg\},
\end{equation}
where $||v||$ denotes the $l_{2}$ norm of the vector $v$. Thus, the $\varepsilon$-averaging time is the smallest time it takes for $\xi(k)$ to be within $\varepsilon$ of $\xi_{ave}\textbf{1}$ with high probability, regardless of the initial value $\xi(0)$.
Let $W$ denote the expected value of $W(0)$ which is the same as $E(W(k))$:  $W=\frac{1}{n}\sum_{i,j}P_{ij}W_{ij}.$
\begin{lemma}[\cite{Boyd2006}]\label{lemma2.1}
	For the randomized consensus algorithm characterized by stochastic matrix $P$, for any initial vector $\xi(0)$, for $k\geq K^{*}(\varepsilon)$
	\begin{equation}
	Pr(\frac{||\xi(t)-\xi_{ave}\textbf{1}||}{||\xi(0)||}\geq \varepsilon)\leq \varepsilon ,
	\end{equation}
	where
	$
	K^{*}(\varepsilon)\triangleq \frac{3\log \varepsilon^{-1}}{\log\lambda_{2}(W)^{-1}}.
$
\end{lemma}

The result in Lemma \ref{lemma2.1} has the following intuitive explanation. For any randomized gossip algorithm with symmetric expectation matrix $E(W)$, the rate of convergence is governed by the second largest eigenvalue $\lambda_{2}(W)$.
\subsection{Centralized Kalman Filtering algorithm Implemented by Randomized Strategy}

Given the system dynamics \eqref{systemdynamic} with observations of $N$ nodes, we derive the distributed implementation of centralized filtering algorithm. Note that the incremental update of our proposed algorithm is similar to the update proposed in \cite{Saber2007(2)}. An important difference in the algorithm is in the consensus step.

	  In \cite{Saber2007(2)}, the authors attempt to reduce the disagreement regarding the state estimates using an \textit{ad hoc} approach by implementing a consensus step right after the estimation step while we adopt a randomized consensus strategy.
	
	The cause of adopting randomized strategy in the communication is the 
	unreliable environment
	where these protocols are implemented: packet losses, collisions and sensor nodes failures are in fact rather common in a sensor network.

The details of proposed centralized Kalman filtering algorithm based on randomized consensus strategy are summarized in \textit{Algorithm 1}.

\begin{algorithm}[htbp]\label{algor1}
	\caption{Centralized Kalman filter based on randomized consensus strategy for sensor $i$}
	\SetAlgoLined
	1: \textbf{Initialization}: $k=0$, $P^{i}_{0|-1}=n\Pi_{0}$, $\hat{x}^{i}_{0|-1}=\textbf{0}$,  consensus iterations $K$\footnotemark[1].\\
	2: \While{ new observation data exists}{
		3: Get observed value $y_{i}(k)$ and measurement information matrix $R^{-1}_{i}$. Compute information vector and matrix:
		\begin{eqnarray*}
			U_{i}(0)=C_{i}R_{i}^{-1}C_{i},
			u_{i}(0)=C_{i}'R^{-1}_{i}y_{i}(k)
		\end{eqnarray*}
		4: Perform the average consensus on $u_{i}$ and $U_{i}$ independently.\\
		\For{$t=1$ to $K$}{
			a) Send $u_{i}(t-1)$ and $U_{i}(t-1)$ to all neighbors $j\in \mathcal{N}_{i} $\\
			b) Receive $u_{j}(t-1)$ and $U_{j}(t-1)$ from all neighbors $j\in \mathcal{N}_{i}$\\
			c) Update
			\begin{equation*}
			U_{i}(t)=W(t)U_{i}(t-1), u_{j}(t)=W(t)u_{j}(t-1)
			\end{equation*}}
		5: Compute the intermediate Kalman estimate of the
		target state:
		\begin{equation*}
		\begin{aligned}
		\quad \quad &(P^{i}_{k})^{-1}= (P^{i}_{k|k-1})^{-1}+U_{i}(K) \\
		\quad \quad &\varphi^{i}_{k}=\hat{x}^{i}_{k|k-1}+P^{i}_{k}[u_{i}(K)-U_{i}(K)\hat{x}^{i}_{k|k-1}]
		\end{aligned}
		\end{equation*}
		6:  Update the state of the local Kalman filter:
		\begin{eqnarray*}
			\hat{x}^{i}_{k+1|k}=A \hat{x}^{i}_{k},
			P^{i}_{k+1|k}=AP^{i}_{k}A'+Q,
	\end{eqnarray*}}
\end{algorithm}
	\begin{lemma}(\textit{cf. Theorem 1 in \cite{Saber2007(2)}}) 
		Assume the nodes of the sensor networks solve two consensus problems that allow them to calculate the average inverse covariance $S$ and the average measurement $q$ at every iteration $k$. Then, every node of the network can calculate the state estimate $\hat{x}_{k}$ at iteration $k$ using the update equations as follows:
		\begin{equation}
		\begin{aligned}
		M_{\mu}&=(P^{-1}_{\mu}+S)^{-1}, \\
		\hat{x}_{k}&=\hat{x}_{k|k-1}+M_{\mu}(q-S\hat{x}_{k|k-1}),\\
		P_{\mu}^{+}&=AM_{\mu}A'+Q_{\mu}, \\
		\hat{x}_{k+1|k}&=A\hat{x}_{k}.
		\end{aligned}
		\end{equation} This gives a distributed estimate identical to the one obtained via a centralized Kalman filtering algorithm.
\end{lemma}

\begin{remark}
	We assume that all nodes know the the number of nodes $n$ in wireless sensor networks or solve a consensus problem to calculate $n$.  Then, the local and central state estimates for all nodes are the same, i.e., $\hat{x}^{c}_{k}=\hat{x}^{i}_{k}$ for all $i$.
\end{remark}
\begin{remark}
	We know that the randomized gossip algorithm reaches the average consensus as time approaches infinity.  In practical application, we define the $\epsilon$-average time which is the smallest time it takes for $x(k)$ to be within $\varepsilon$ of $x_{ave}\textbf{1}$ with high probability. In other words,  if consensus iterations $K \geq  \frac{3\log \varepsilon^{-1}}{\log\lambda_{2}(W)^{-1}},$ the system state $x(k)$ becomes within $\varepsilon$ of $x_{ave}\textbf{1}$.
\end{remark}

	\footnotetext[1]{ When $K$ is larger than $K^{*}$, the system state can become within $\epsilon$ of the consensus value with high probability.}

	This network of consensus scheme for centralized Kalman filtering algorithm is able to collaboratively provide an identical performance to the estimates obtained by a central Kalman filter given that all nodes agree on the two central sums.
	Randomized gossip algorithm can approximate these sums and gives an approximate distributed Kalman filter for sensor networks.  However, \textit{Algorithm 1} does not solve the distributed Kalman filtering problem. So far, we have shown that if two dynamic consensus problems in $S$ and $q$ are solved, then, a centralized Kalman filtering algorithm can be solved in a distributed way.
This paper will devote to
developing a novel randomized consensus based distributed Kalman filtering algorithm.

\subsection{Problems of Interest}

In our work, due to the limited single-sensor energy, computational ability, and communication capability, a large number of sensor nodes are commonly used in a wide region to estimate the same system state.  Each sensor, through communication with its neighbors to exchange information, simultaneously estimates the global performance of the system
and updates its own estimates to improve the performance. Eventually, all nodes could reach a consensus agreement about the values of their estimates of the state. The main problems of interest are summarized as follows:
\begin{enumerate}
	\item How to design a distributed estimation strategy that is adaptive to network topology changes?
	\item What conditions can guarantee the convergence of the proposed distributed estimation algorithm?
	\item How to maximize the estimation performance under limited energy budget?
\end{enumerate}

The detailed formulations and solutions to these problems are presented in the following section.

%
%
%
%

\section{Randomized Consensus based Distributed Kalman Filtering}\label{sec:RGDF}
    In this section, we propose a novel approach to distributed Kalman filtering which depends on the estimation communication among neighboring nodes based on randomized gossip algorithm. This approach  is referred to as randomized gossip based distributed Kalman filtering algorithm.
    Before presenting this distributed Kalman filtering iteration, it is essential to introduce a more primitive noncooperative decentralized Kalman filtering algorithm that forms the basis of our algorithm.
    \subsection {Decentralized Estimation Process}
    Decentralized Kalman filtering algorithm have drawn a lot of
    interest during the past few decades in
    WSNs due to that they do not need a centralized processing
    station \cite{Rao1993}, \cite{Sanders1974}.
      \begin{figure}
      	\centering
      	\includegraphics[width=1\linewidth]{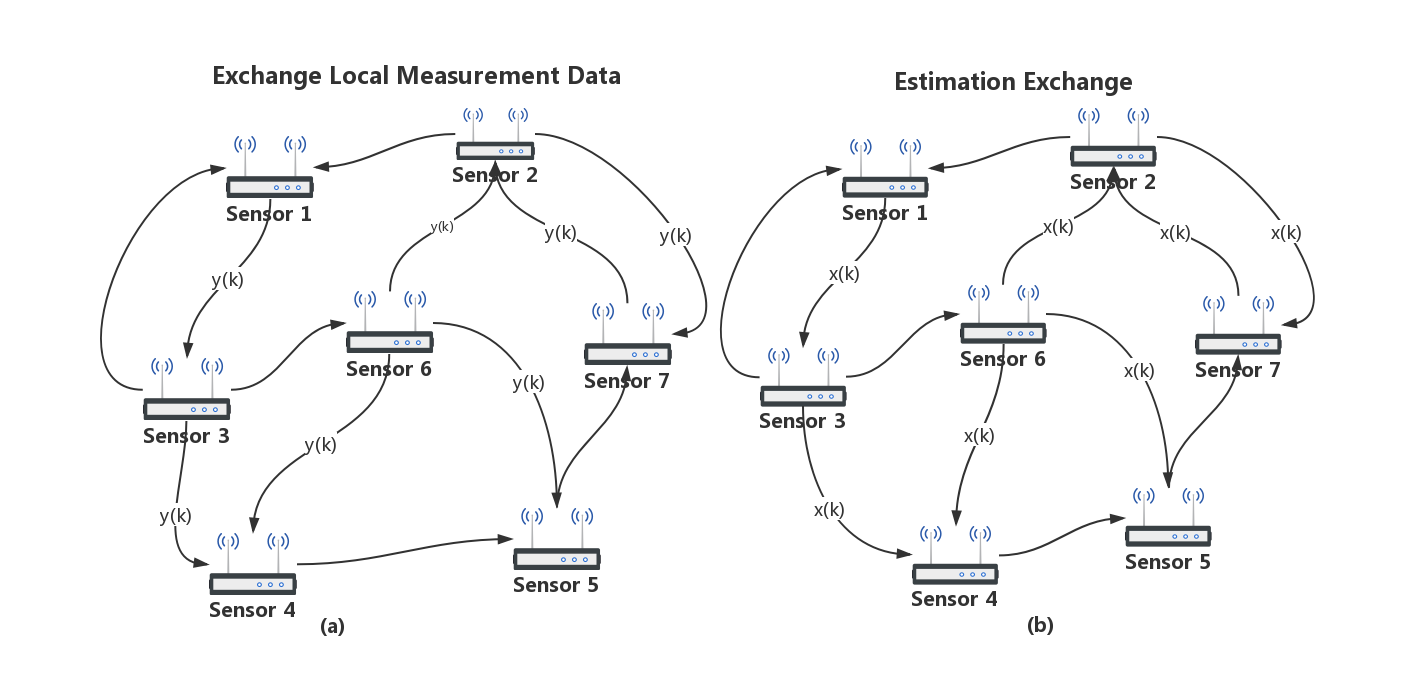}\\
      	\caption{A WSN exchanging information. (a) Measurement exchange; (b) Estimation exchange.}
      \end{figure}

    Assume that sensor node $i$ of the sensor network can exchange its observed value $y_{i}$, covariance information matrix $R_{i}$, and output matrix $C_{i}$ with its neighboring nodes $\mathcal{N}_{i}$.
    At time $k$, all the sensor nodes first locally predict the state $x_{k}$. Then they transmit their local observed value to and receive these from their neighboring sensor nodes through available channels.
    After the data communication, they update their local estimation.
    For the $i$th sensor node, denote $\hat{x}^{i}_{k|k-1}$ as the priori estimate of $x(k)$, which is the predicted state estimate, and $\hat{x}^{i}_{k}$ as the posteriori estimate of $x(k)$ after updating  the observed value
    both taken by itself locally and sent by the other neighboring sensor nodes. Further denote $P^{i}_{k|k-1}$ and $P^{i}_{k}$ as the estimation error
    covariance matrices of $\hat{x}^{i}_{k|k-1}$ and $\hat{x}^{i}_{k}$, respectively.
     Computation of the aforementioned quantities are given as follows:
     \begin{itemize}
    	 \item   At time $k$, sensor node $i$ first computes $\hat{x}^{i}_{k|k-1}$ and $P^{i}_{k|k-1}$ based on the following equations
    	\begin{align*}
    	\hat{x}^{i}_{k|k-1}=A \hat{x}^{i}_{k-1},
    	P^{i}_{k|k-1}=AP^{i}_{k-1}A'+Q,
    	\end{align*}
    	where the recursion starts from $\hat{x}^{i}_{0}=0$ and $P^{i}_{0}=\Pi_{0}$.
    	\item  After the local observed value $y^{i}_{k}$ is acquired, sensor node $i$ transmits $y^{i}_{k}$ to their neighboring sensor nodes and receives the data from neighboring nodes by available edges. Then the sensor nodes first do the fusion of information. For sensor node $i$, define
    	\begin{align*}
    	S_{i} = \sum_{l \in \mathcal{N}_{i} }C_{l}R_{l}^{-1}C_{l}', \quad
    	q^{i}_{k} = \sum_{l\in \mathcal{N}_{i}}C_{l}'R^{-1}_{l}y_{l}(k).
    	\end{align*}
    	\item Then for node $i$, the incremental update is computed as follows:
    	\begin{align*}
    	(P^{i}_{k})^{-1}&= (P^{i}_{k|k-1})^{-1}+S_{i}, \\
    	\hat{x}^{i}_{k}&= \hat{x}^{i}_{k|k-1}+P^{i}_{k}[q^{i}_{k}-S_{i}\hat{x}^{i}_{k|k-1}].
    	\end{align*}
    \end{itemize}
%
    According to the standard Kalman filter, we know that $P^{i}_{k}$ converges to a steady-state value exponentially fast \cite{Yang2013}. Define
    \begin{align}
    P_{i}\triangleq \lim_{k\to \infty} P^{i}_{k}.
    \end{align}
    The computation of equations above shows that $P_{i}$ depends on the values of adjacent matrix $\Gamma$.

    Should one avoid implementing any consensus, i.e., without further information communication regarding state estimations? The answer to this problem is rather simple. Actually, one can adopt local decentralized Kalman filtering which acts as a basis performance standard for distributed Kalman filtering algorithms. Intuitively, local decentralized Kalman filtering algorithm does not behave well due to the fact that a minority of nodes and their neighbors make poor observations due to environmental or geometric factors.

    In local decentralized Kalman filtering algorithm, we assume that no nodes apart from its neighboring nodes
    $\mathcal{N}_{i}$ can transmit the information data straight to node $i$. The case that its neighboring nodes transmit the information data from non-neighboring nodes to node $i$ is forbidden.
    Consequently, sensor node $i$ can use a central Kalman filter that only utilize the observed values and output matrices of  neighboring nodes. This results in the following primitive decentralized Kalman filtering iterations without a consensus on state estimates.

    The optimal local state estimation above is generated by incrementally incorporating estimates and data sequentially from the neighborhoods. The iterations calculate the optimal estimate for every neighborhood only. In \cite{Saber2007(2)}, the authors proposed the novel distributed Kalman filtering algorithms for WSNs that have a wide range of applications.
   \begin{algorithm}
 	\caption{Randomized consensus based distributed Kalman filtering algorithm (distributed Kalman filtering algorithm with a randomized gossiping step on estimates).}
 	\label{algorithm}
 	\SetAlgoLined
 	1: \textbf{Initialization}: $P_{0|-1}=\Pi_{0}$, $\hat{x}_{0|-1}=0$, $k=0$, consensus iterations $K$\footnotemark[2].  \\
 	2: At every time instant $k$:	\\
 	\For{$i=1 \to n$ }{
 		a) Get the observed value $y_{i}(k)$. Compute the information vector and matrix
 		\begin{eqnarray*}
 			u_{i}=C_{i}R_{i}^{-1}C_{i},\quad
 			U_{i}=C_{i}'R^{-1}_{i}y_{i}(k);
 		\end{eqnarray*} \\
 		b) Broadcast information data containing $u_{i}$ and $U_{i}$ to neighboring nodes.  \\
 		c) Receive information data from neighbors $l \in N_{i} $. Locally aggregate observed data and covariance matrices:
 		\begin{eqnarray*}
 			S_{i} = \sum_{l \in N_{i} }C_{l}R_{l}^{-1}C_{l}' ,\quad
 			q^{i}_{k} = \sum_{l\in N_{i}}C_{l}'R^{-1}_{l}y_{l}(k),
 		\end{eqnarray*}
 		d) Compute the intermediate Kalman estimate of the target state:
 		\begin{eqnarray*}
 			(P^{i}_{k})^{-1}&=& (P^{i}_{k|k-1})^{-1}+S_{i} \\
 			\varphi^{i}_{k}&=&\hat{x}^{i}_{k|k-1}+P^{i}_{k}[q^{i}_{k}-S_{i}\hat{x}^{i}_{k|k-1}]
 		\end{eqnarray*}
 		e) Estimate the target state after a randomized gossiping step: we first define
 		\begin{align*}
 		\hat{x}_{k}=[(\hat{x}^{1}_{k})',(\hat{x}^{2'}_{k})',(\hat{x}^{3}_{k})',...,(\hat{x}^{n}_{k})']'
 		\end{align*} and reduce the estimation disagreement of different sensor nodes:\\
 		\For{$t=1 \to K$}{
 			\begin{eqnarray*}
 				& \hat{x}_{k}=(W(t) \otimes I_{m})[(\varphi^{1}_{k})',(\varphi^{2}_{k})',...,(\varphi^{n}_{k})']',\\
 				& [(\varphi^{1}_{k})', (\varphi^{2}_{k})'...,(\varphi^{n}_{k})']'=[(\hat{x}^{1}_{k})',...,(\hat{x}^{n}_{k})']'.
 			\end{eqnarray*}
 	}}
 	f) Update the state of the local Kalman filter:
 	\begin{eqnarray*}
 		\hat{x}^{i}_{k+1|k}=A \hat{x}^{i}_{k},\quad
 		P^{i}_{k+1|k}=AP^{i}_{k}A'+Q.
 	\end{eqnarray*}
 \end{algorithm}

     The main difference between the above iteration and  \cite{Saber2007(2)} are summarized as follows.  \cite{Saber2007(2)} attempts to diminish the disagreement with regard to the state estimations in local Kalman filtering adopting an \emph{ad hoc} approach by implementing a consensus step right after the estimation step
    \begin{align}\hat{\phi}^{i}_{k}=\hat{x}^{i}_{k}+\epsilon\sum_{j \in N_{i}}(\hat{x}^{j}_{k}-\hat{x}^{i}_{k}).\end{align}
    This is equivalent to moving toward the average intermediate estimate of the neighboring nodes.
    Based on the local Kalman filtering and randomized gossiping algorithm, we derive the distributed Kalman filtering algorithm updated by adding a random consensus step between the Kalman filtering updates.
 The randomized gossiping step is an attempt to achieve the global state consensus by local node interaction.
 The detailed realization of the randomized gossip based distributed Kalman filtering algorithm is as shown in \textit{Algorithm 2}.
 \footnotetext[2]{When $K$ is larger than $K^{*}$ which is defined in \textit{Remark 1}, the system state can become within $\epsilon$ of the consensus value with high probability.}
 The objective in the above algorithm is to diminish the estimation disagreement of different nodes.
 The proposed randomized consensus algorithm exchanges information and computes in an arbitrary network. Besides, the proposed algorithm  distributes the computational burden by communicating with a randomly chosen neighbor. It can be shown that the proposed algorithm has a better performance than local noncooperative decentralized Kalman filtering algorithm. The convergence and performance analysis are provided below.
\begin{remark}
The neighboring nodes of sensors are exclusively determined by the network topology at the sampling instant, while during sensor communication it is a set with random neighbors selected according to the randomized algorithm. With neighboring node being deterministic in the former case, the filtering algorithm is guaranteed to converge due to the connectivity condition indicated by $N_{i}$; in contrast, the random of neighboring node is exploited to improve the performance in the latter case.
\end{remark}

Although the consensus based Kalman filtering algorithm is highly interesting, the proof of convergence and performance analysis are relatively difficult so far. It is still an open issue. Our own work is in fact motivated by these results extended from Saber \cite{Saber2007(2)}. The main difference is that we use a different consensus algorithm in the paper and provide a convergence proof as follows.
\subsection{Convergence Analysis of Proposed Distributed Filtering Algorithm}
	In this section, we analyze the mean-square error performance of the proposed algorithm.
    Let $\widetilde{\varphi}^{i}_{k}=x_{k}-\varphi^{i}_{k}$ denote the estimation error at the end of the incremental update.
    Denote $\widetilde{x}^{i}_{k|k-1}=x_{k}-\hat{x}^{i}_{k|k-1}$ and  $\widetilde{x}^{i}_{k|k}=x_{k}-\hat{x}^{i}_{k|k}$ as the estimation error at the  sensor node $i$. Furthermore, let
    \begin{eqnarray*}
    	\widetilde{x}_{k|k}&=&[(\widetilde{x}^{1}_{k|k})',(\widetilde{x}^{2}_{k|k})',...,(\widetilde{x}^{n}_{k|k})']',\\\
    	\widetilde{\varphi}_{k}&=&[(\widetilde{\varphi}^{1}_{k})',(\widetilde{\varphi}^{2}_{k})',...,(\widetilde{\varphi}^{n}_{k})']'.
    \end{eqnarray*} It follows that
    \begin{eqnarray*}
      \widetilde{\varphi}^{i}_{k} &=& x_{k}-\varphi^{i}_{k} \\
       &=& x_{k}-\hat{x}^{i}_{k|k-1}-P^{i}_{k}\sum_{l \in \mathcal{N}_{i}}C_{l}'R^{-1}_{l}[y^{l}_{k}-C_{l}\hat{x}^{l}_{k|k-1}]\\
        &=& \widetilde{x}^{i}_{k|k-1}- P^{i}_{k}\sum_{l \in \mathcal{N}_{i}}C_{l}'R^{-1}_{l}[C_{l}\widetilde{x}^{i}_{k|k-1}+v_{l}]\\
       &=& P^{i}_{k}[(P^{i}_{k})^{-1}-S_{i}]\widetilde{x}^{i}_{k|k-1}-P^{i}_{k}\sum_{l \in \mathcal{N}_{i}}C_{l}'R^{-1}_{l}v_{l}\\
       &=& P^{i}_{k}(P^{i}_{k|k-1})^{-1}\widetilde{x}^{i}_{k|k-1}-P^{i}_{k|k}\sum_{l \in \mathcal{N}_{i}}C_{l}'R^{-1}_{l}v_{l}.
    \end{eqnarray*}
    We also have
    \begin{align}
        \widetilde{x}^{i}_{k|k-1}=A \widetilde{x}^{i}_{k-1|k-1}+w(k).
    \end{align}
    Combining the above two equations into the consensus step, we obtain
    \begin{eqnarray}
    \widetilde{x}_{k|k}=(W(k) \otimes I_{m}) \widetilde{\varphi}_{k}.
    \end{eqnarray}
Define $v_{k}=[(v^{1}_{k})',(v^{2}_{k})',(v^{3}_{k}),...,(v^{n}_{k})']'$ and $\mathcal{C}=diag\{C_{1},C_{2},C_{3}...C_{n}\}$. Besides, the priori and posteriori estimation error matrices of $n$ sensors are defined in a general form a
    \begin{equation}
    \begin{aligned}
    \mathcal{P}_{k|k-1}&=diag\{P^{1}_{k|k-1},...,P^{n}_{k|k-1}\},\\
    \mathcal{P}_{k}&=diag\{P^{1}_{k},...,P^{n}_{k}\}.
    \end{aligned}
    \end{equation}

    We also define that
    \begin{eqnarray*}
        \mathcal{G}&=&[(C_{1}R^{-1}_{1}v_{1})',(C_{2}R^{-1}_{2}v_{1})',...,(C_{n}R^{-1}_{n}v_{n})']',\\
        \mathcal{H}&=&\begin{bmatrix}
            P^{1}_{k}(P^{1}_{k|k-1})^{-1}[A \widetilde{x}^{1}_{k-1|k-1}+w(k)] & \\
            P^{2}_{k}(P^{2}_{k|k-1})^{-1}[A \widetilde{x}^{2}_{k-1|k-1}+w(k)] & \\
            \vdots & \\
            P^{n}_{k}(P^{n}_{k|k-1})^{-1}[A \widetilde{x}^{n}_{k-1|k-1}+w(k)]
            \end{bmatrix}.
    \end{eqnarray*}
   The extended matrices may be defined as
    \begin{align}
      \mathcal{W} \triangleq W(k)\otimes I_{m}.
    \end{align}
    We further define a link matrix $\Gamma$ as follows:
    \begin{align}
         &[\Gamma]_{l,k} = \left\{ \begin{array}{ll}
    1 ,& \textrm{if $l \in \mathcal{N}_{k}$}, \\
    0 ,& \textrm{elsewhere},\\
    \end{array} \right.
    \end{align}and its extended matrix
    \begin{align}
    \mathcal{L}\triangleq \Gamma \otimes I_{m},
    \end{align} where $\otimes$ denotes the Kronecker product. Observe that
    \begin{align}
        \widetilde{x}_{k|k}=\mathcal{W}'\mathcal{H}-\mathcal{W}'\mathcal{P}_{k}\mathcal{L}'\mathcal{G}
    \end{align} or equivalently
    \begin{equation}
    \begin{aligned}\label{24}
    \widetilde{x}_{k|k}=&\mathcal{W}\mathcal{P}_{k}[\mathcal{P}_{k|k-1}^{-1}(\mathbf{1} \otimes A)\widetilde{x}_{k-1|k-1}
    +\mathcal{P}_{k|k-1}(I   \\
    & \otimes w(k))-\mathcal{L}'\mathcal{C}R^{-1}v_{k}].
    \end{aligned}
    \end{equation}

    \begin{remark}
    	Taking expectations of both sides of equation \eqref{24}, we obtain the following recursion for the expectation of the estimate of the randomized gossip based Kalman filtering algorithm
    	\begin{equation}\label{eee}
    	E[\widetilde{x}_{k|k}]=E[\mathcal{W}\mathcal{P}_{k}[\mathcal{P}_{k|k-1}^{-1}(\mathbf{1} \otimes A)E[\widetilde{x}_{k-1|k-1}]].
    	\end{equation}
    	Since $E[\widetilde{x}_{0|-1}]=0$ and $E[\widetilde{x}_{0|0}]=0$, we conclude from \eqref{eee} that the proposed randomized consensus Kalman filtering estimation is unbiased.
    \end{remark}

   We provide the mean-square performance analysis in the following. Let $\mathcal{A}_{k}= \mathcal{P}_{k}\mathcal{P}_{k|k-1}^{-1}(\mathbf{1} \otimes A)$, $\mathcal{B}_{k}= \mathcal{P}_{k}\mathcal{P}_{k|k-1}^{-1},$ and
   $\mathcal{D}_{k}=\mathcal{P}_{k}\mathcal{L}'\mathcal{C}R^{-1}$. Equation \eqref{24} can be rewritten in a more compact form as
    \begin{align}\label{50}
        \widetilde{x}_{k|k}=\mathcal{W}\mathcal{A}_{k}\widetilde{x}_{k-1|k-1}+\mathcal{W}\mathcal{B}_{k}(\mathbf{1} \otimes w(k))- \mathcal{W}\mathcal{D}_{k}v_{k}.
    \end{align}

\begin{assumption}
	The sensor communications are much faster than measurements so that consensus can be reached during two consecutive measurements.
\end{assumption}

\begin{assumption}
	The local decentralized noncooperative Kalman filtering of each node that uses the data from neighborhoods  converges to a steady value as time approaches infinity, i.e., $\lim\limits_{k\rightarrow{\infty}} P^{i}_{k|k-1}={P}_{i}^{-}$ and $\lim\limits_{k\rightarrow{\infty}}P^{i}_{k}={P}_{i}$, for~$\forall i\in \{1,2,...,n\}$ (see \cite{Kailath2000} for conditions on Kalman filter convergence).
\end{assumption}

Note that Assumption $1$ was used in \cite{Saber2007,Saber2005a}.
Under the  Assumption $2$, the matrices $\mathcal{A}, \mathcal{B}, \mathcal{D}$ also converge to the steady-state, and their steady-state values are given by
\begin{eqnarray*}
    \mathcal{P} &\triangleq& \lim\limits_{n\rightarrow{\infty}} \mathcal{P}_{k}=diag\{P_{1},P_{2},...P_{n}\},\\
    \mathcal{P}^{-} &\triangleq& \lim\limits_{n\rightarrow{\infty}} \mathcal{P}_{k|k-1}=diag\{P_{1}^{-},P_{2}^{-},...P_{n}^{-}\},\\
    \bar{\mathcal{A}} &\triangleq& \lim\limits_{k\rightarrow{\infty}} \mathcal{A}_{k}=\mathcal{P}(\mathcal{P}^{-})^{-1}(I \otimes A),\\
    \bar{\mathcal{B}} &\triangleq& \lim\limits_{k\rightarrow{\infty}} \mathcal{B}_{k}=\mathcal{P}(\mathcal{P}^{-})^{-1}(I \otimes Q),\\
    \bar{\mathcal{D}} &\triangleq& \lim\limits_{k\rightarrow{\infty}} \mathcal{D}_{k}=\mathcal{P}\mathcal{L}'\mathcal{C}'R^{-1}.
\end{eqnarray*}
    Let $\mathcal{P}_{\widetilde{x},k}=E\{\widetilde{x}_{k|k}\widetilde{x}_{k|k}'\}$ denote the covariance matrix of the proposed estimation algorithm. When time $k$ is sufficiently large, based on equation \eqref{50}, the whiteness of noise on the state and the observed value, we obtain
    \begin{equation}
    \begin{aligned}
    \mathcal{P}_{\widetilde{x},k}&=\mathcal{W}\mathcal{\bar{A}}\mathcal{P}_{\widetilde{x},k-1}\mathcal{\bar{A}}'\mathcal{W}'
    +\mathcal{W}\mathcal{\bar{B}}(\mathbf{1}\mathbf{1}'\otimes Q)\mathcal{\bar{B}}'\mathcal{W}'
    \\
    &+\mathcal{W}\mathcal{\bar{D}}R\mathcal{\bar{D}}'\mathcal{W}'.
    \end{aligned}
    \end{equation}
    When it is not involved in a randomized gossiping step, the sensor node does not exchange estimation information with neighbors, it is consistent with the local noncooperative decentralized Kalman filtering algorithm.  The estimation error covariance matrix $\mathcal{P}_{k}$
    converges to $\mathcal{P}$ as $k$ approaches infinity, where $\mathcal{P}$ is the steady state estimation error covariance matrix in decentralized Kalman filtering algorithm. Before providing a mean-square performance analysis, we introduce the following lemma.


\begin{lemma}\label{lemma1}
	$Tr(P)\geqslant Tr(WPW')$, where $W$ is a symmetric stochastic matrix and $P$ is a symmetric positive-definite matrix.
\end{lemma}
\begin{IEEEproof}
	Firstly, it is straightforward to have the following property \cite{Wilkinson2013}:
	$Tr(ABC)=Tr(CAB)=Tr(CBA)$. Due to the fact that
	$W$ and $P$ are all symmetric matrices, it is easy to obtain that $Tr(WPW')=Tr(W'WP)$. In order to demonstrate that $Tr(P)\geqslant Tr(WPW')$, it is equivalent to prove that $Tr(P)\geqslant Tr(W'WP)$, i.e., $Tr(P-W'WP)=Tr((I-W'W)P)\geqslant 0$. Owing to the fact that $W$ is a doubly stochastic matrix, $W'W$ is also a stochastic matrix and the maximal eigenvalue is equal to 1. Hence, $I-W'W$ is a positive semi-definite matrix due to the fact that $x(I-W'W)x' \geqslant 0$ for any vector $x$. By using Cholesky factorization, $P$ can
	be factorized as $P=RR'$. We can see that $Tr((I-W'W)P)=Tr((I-W'W)RR')=Tr(R'(I-W'W)R)$, and $R'(I-W'W)R$ is a positive semi-definite matrix because $x'R'(I-W'W)Rx \geqslant 0$ for any vector $x$. Consequently, $Tr(R'(I-W'W)R)\geqslant 0$. This proves that $Tr(P)\geqslant Tr(W'WP)$ and completes the proof of lemma \ref{lemma1}.
	
	The stability and convergence analysis of the proposed algorithm are summarized in the following theorem.
\end{IEEEproof}
    \begin{theorem}
    The estimation error covariance of our proposed algorithm is defined as $\mathcal{P}_{\widetilde{x},k}=E\{\widetilde{x}_{k|k}\widetilde{x}_{k|k}'\}$ and
    $\mathcal{P}_{\widetilde{x},k}$ follows the iteration:
    \begin{eqnarray}
        &E[\mathcal{P}_{\widetilde{x},k}]=T(E[\mathcal{P}_{\widetilde{x},k-1}]):R^{nm \times nm}\rightarrow R^{nm \times nm} \notag,
        \end{eqnarray}
        where
       $T(\mathcal{P}_{\widetilde{x},k})\triangleq E[\mathcal{W}\mathcal{\bar{A}}\mathcal{P}_{\widetilde{x},k-1}\mathcal{\bar{A}}'\mathcal{W}'
    +\mathcal{W}\mathcal{\bar{B}}(\mathbf{1}\mathbf{1}'\otimes Q)\mathcal{\bar{B}}'\mathcal{W}'
    +\mathcal{W}\mathcal{\bar{D}}R\mathcal{\bar{D}}'\mathcal{W}'] \notag
    $
    and $\mathcal{P}_{0}=\mathcal{P}_{\widetilde{x},0} \geq 0$. Then the expectation of estimation error covariance matrix $E[\mathcal{P}_{\widetilde{x},k}]$ converges exponentially to a unique fixed point $\mathcal{\bar{P}}_{\widetilde{x}}$ of the mapping $T$.
    \begin{IEEEproof}
	See the proof in Appendix A.
    \end{IEEEproof}
    \end{theorem}

Next, we show that our proposed algorithm achieves better performance than
decentralized Kalman filtering algorithm under a sufficient condition in the following theorem.
    \begin{theorem}
    	Let $\mathcal{P}_{k}$ be the error covariance matrix of noncooperative decentralized Kalman filtering algorithm such that $\mathcal{P}_{k}$ converges to the unique steady state $\mathcal{P}$ for any initial condition.
  Suppose that $\mathcal{\bar{A}}=\mathcal{P}(\mathcal{P}^{-})^{-1}(I \otimes A)$ is an orthogonal matrix, then $Tr(E[\mathcal{P}_{\widetilde{x},k}]) \leqslant Tr(\mathcal{P}_{k})$, $\forall k\in R$.
    \end{theorem}
    \begin{IEEEproof}
   The proof is shown in Appendix B.
   \end{IEEEproof}
\begin{remark}
	To compute the randomized gossip based Kalman filtering algorithm, node $i$ needs to have knowledge of $ (P^{i}_{k|k-1})^{-1}$. In general, computation of $ (P^{i}_{k|k-1})^{-1}$ requires the knowledge of the entire covariance matrix (i.e., the prior covariance of each node and the prior cross-covariances between each pair of nodes). However, computing $(\mathcal{P}^{-})^{-1}$ at every time step at each node in a distributed framework is unrealistic as it would require too much information. When the prior state estimates across the nodes are uncorrelated to each other\cite{Kamal2013}, in this case the $(\mathcal{P}^{-})^{-1}$ can be computed at each node using only a node's own prior covariance matrix (which is of great practical importance).
\end{remark}

In the next section, we will provide a sub-optimal sensor
scheduling scheme for distributed estimation subject to limited power.
\section{Optimal Sensor Connection Scheme}\label{sec:Problem statement}
    Some resources such as battery power or channel bandwidth are consumed when the sensors transmit information in WSNs. In this section, we aim to minimize the estimation error while guaranteeing that the specified resource consumption is within a budget.
    We consider the average steady-state estimation
    error of the $n$ sensor nodes:
    \begin{align}
        J(\Gamma)=\frac{1}{n}\sum^{n}_{i=1}Tr(E[\mathcal{{P}}_{\widetilde {x}, i}]),
    \end{align}
    where $\mathcal{{P}}_{\widetilde {x}, i}$ is the steady state estimation error of sensor $i$ when adopting randomized gossip consensus based distributed Kalman filtering algorithm.

    We consider the following power constraint for sensor $i$:
    \begin{align}
        \sum_{j\in N_{i}}c_{ij}\gamma_{ij}+\frac{1}{n}P_{ij}c_{ij} \leq \delta_{i},
    \end{align}
    where $\delta_{i}$ is a given constant quantifying the power budget of sensor $i$ imposed at each time instant, and $c_{ij}$ is a parameter indicating the communication cost on transmitting data from sensor $i$ to sensor $j$. And node $i$ is active with probability $\frac{1}{n}$, it will contact one neighbor $j$ with probability $P_{ij}$. In practice, the $c_{ij}'s$ can be different due to different location of the sensors in a large range. We design the optimal sensor schedule while adopting proposed algorithm as:
    \begin{problem}\label{pro1}
        \begin{eqnarray}\label{problem1}
        &min& J(\Gamma) \notag\\
        &s.t.& \sum_{j\in N_{i}}c_{ij}\gamma_{ij}+\frac{1}{n}P_{ij}c_{ij} \leq \delta_{i},\\
        &for& all \quad i=1,2,...,n, \gamma_{ij}=0,1.\notag
        \end{eqnarray}
    \end{problem}

    In section \ref{sec:RGDF}, we have proved that the upper bound of the expected error covariance of our proposed algorithm converges to a unique limit. This shows that the steady state error covariance of our algorithm exists. We turn to tackle Problem \ref{pro1} formulated in \eqref{problem1}.
    To solve Problem \ref{pro1}, we transform the implicit form of $J(\Gamma)$ with respect to the optimization variable into explicit one.
    Assume that $\gamma_{ij}$ and $\gamma_{ji}$ are independent, we can decompose the Problem \ref{pro1} into $n$ independent  optimization problems as follows:
    \begin{problem}\label{pro2}
        \begin{eqnarray}
        &min& \frac{1}{n}Tr(E[\mathcal{{P}}_{\widetilde {x}, i}]) \notag\\
        &s.t.& \sum_{j\in N_{i}}c_{ij}\gamma_{ij}+\frac{1}{n}P_{ij}c_{ij} \leq \delta_{i},\\
        &for& \gamma_{ij}=0,1.\notag
        \end{eqnarray}
    \end{problem}

    Since the explicit form of the state expected error covariance $E[\mathcal{{P}}_{\widetilde {x}, i}]$ is difficult to obtain due to its randomness, we replace the objective function in Problem \ref{pro2} by the upper bound of the steady-state error covariance $\mathcal{P}_{i}$. 
    We relax Problem \ref{pro2} as the following optimization problem:
     \begin{problem}\label{pro3}
        \begin{eqnarray}
        &min& \frac{1}{n}Tr(\mathcal{P}_{i}) \notag\\
        &s.t.& \sum_{j\in N_{i}}c_{ij}\gamma_{ij}+\frac{1}{n}P_{ij}c_{ij} \leq \delta_{i},\\
        &for& \gamma_{ij}=0,1.\notag
        \end{eqnarray}
    \end{problem}

    One of the difficulties to solve the Problem lies in the implicit form of $\mathcal{P}_{i}$ with respect to the optimization variables. We address this difficulty in the following part.
    Using the Cholesky factorization, $S_{i}$ can be factorized as
    $S_{i}=H_{i}'H_{i}$. Let $H\triangleq(H_{1}',H_{2}',...,H_{n}')'$. Define $\Xi\triangleq diag\{\gamma_{i1}I_{p_{1}},\gamma_{i2}I_{p_{2}},...,\gamma_{in}I_{p_{n}}\}$, where $I_{p_{i}}$ is the identity matrix with order $p_{i}$, i.e., the order of $y_{i}(k)$. We further define the operator $\hat{g}(X;\Xi)$ as
    \begin{align}
        \hat{g}(X;\Xi)\triangleq ([h(X)]^{-1}+H'\Xi H)^{-1},
    \end{align}
    where $h(X)=AXA'+Q$.
    Notice that $\mathcal{P}_{i}$ satisfies $\mathcal{P}_{i}=\hat{g}(\mathcal{P}_{i};\Xi)$. Then Problem \ref{pro3} is equivalent to the following problem:
    \begin{problem}\label{pro4}
        \begin{eqnarray}
            &min_{\Xi,X}& \frac{1}{n}Tr(X) \notag\\
            &s.t.& X \geqslant \hat{g}(X;\Xi), \notag\\
            && \sum_{j\in N_{i}}c_{ij}\gamma_{ij}+\frac{1}{n}P_{ij}c_{ij} \leq \delta_{i}, \\
            && \gamma_{ij}=0,1.\notag
        \end{eqnarray}
    \end{problem}

    Problem \ref{pro4} is still not solvable using any efficient numerical algorithm since the feasible domains given by $X \leqslant \hat{g}(X;\Xi)$ and $\gamma_{ij}=0,1$ are not convex. For the former inequality, one \cite{Sinopoli2004} has the following result.
    \begin{lemma} If $(A,C_{i})$ is detectable and $(A,\sqrt{Q})$ is controllable, the following statements are equivalent: \label{lemma2}
    \begin{enumerate}
      \item $\exists X $ such that $X \geqslant \hat{g}(X,\Xi). $
      \item $\exists Z,Y$ such that \\\
      $\left[
        \begin{matrix}
          Y & (Y -Z C_{i})\hat{A} & Y- Z C_{i} & Z \\
          \hat{A}'(Y-C_{i}'Z') & Y & 0 & 0 \\
          Y-C_{i}'Z' & 0 & Q^{-1} & 0 \\
          Z' & 0 & 0 & \Xi\\
        \end{matrix}
      \right]\geqslant 0.$ Moreover, for $Y$ satisfying the inequality in $2)$, $X=Y^{-1}$ is a solution to the inequality in $1)$. It is also true conversely.
    \end{enumerate}
    \end{lemma}
    \begin{IEEEproof}
    The proof is similar to that of Theorem 5 in \cite{Sinopoli2004}.
    \end{IEEEproof}

	Recalling the properties of Lemma \ref{lemma2}, it can be seen that Problem \ref{pro4} is equivalent to the following one:
    \begin{problem}\label{pro5}
    \begin{equation*}
    \begin{aligned}
       &min_{\Xi,X,Y,Z} \frac{1}{n}Tr(X)   \\
       &s.t.  \left[ \begin{matrix} X & I \\ I & Y \end{matrix} \right]\geqslant 0 ,&  \\
       & \left[ \begin{matrix}Y & (Y -Z C_{i})\hat{A} & Y- Z C_{i} & Z \\ \hat{A}'(Y-C_{i}'Z') & Y & 0 & 0 \\ Y-C_{i}'Z' & 0 & Q^{-1} &  0 \\ Z' & 0 & 0 & \Xi \\ \end{matrix}\right] \geqslant 0 , \\
       &\sum_{j\in N_{i}}c_{ij}\gamma_{ij}+\frac{1}{n}P_{ij}c_{ij} \leq \delta_{i},\gamma_{ij}=0,1.
       \end{aligned}
   \end{equation*}
    \end{problem}

    Since the feasible domains of $\gamma_{ij}$ is discrete, Problem \ref{pro5} is a Boolean-convex problem which is a common issue when applying numerical methods. A relaxation on the feasible domains is often used to obtain a convex problem \cite{Joshi2009}.
    \begin{remark}
        Considering the problem which is the same as Problem \ref{pro5} except that the constraint $\gamma_{ij}=0,1$ is replaced by $0\leqslant \gamma_{ij}\leqslant 1$. Denote the solution by $\Gamma^{+}$ and the optimal value by $J^{+}$. Although it is not equivalent to the original problem, the optimal objective value of this relaxed problem is clearly seen to be a lower bound of Problem \ref{pro5}, i.e.,  $J^{*} \geqslant J^{+}$. The elements of $\Gamma^{+}$ may be fractional. We use $\Gamma^{+}$ to obtain a feasible solution to Problem \ref{pro5} denoted as
        $\Gamma^{f}$, which chooses the first $d$ largest element in each row of matrix $\Gamma^{+}$. Although this solution is a relaxed one, the
        discretized solution $\Gamma^{f}$ should be close to or may even coincide with the optimal solution $\Gamma^{*}$.
    \end{remark}
\section{Simulation Example} \label{sec:simulation}
	We present below one example to illustrate our main results.
	\begin{example}
	\begin{figure}
		\centering
		\includegraphics[width=0.5\textwidth]{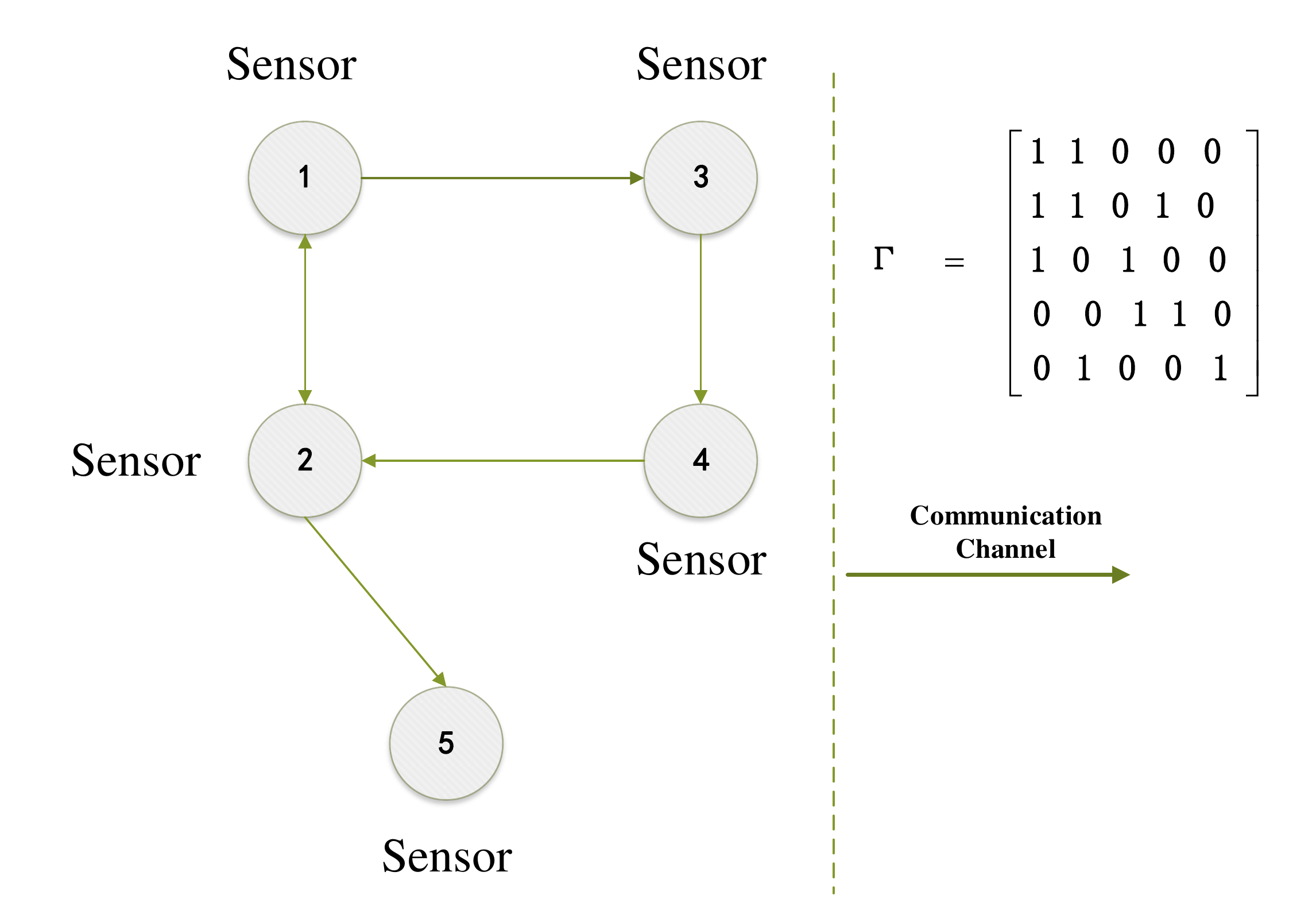}\\
		\caption{The network topology}\label{topology}
	\end{figure}
    To illustrate the better performance of our proposed algorithm compared with the distributed solution in \cite{Saber2007(2),Cattivelli2010}, we present a simulation example in this section and visualize the results in the following.
    Consider a WSN composed of $n=5$ sensors with the network topology in Fig.~2. The system parameters and the adjacency matrix $\Gamma$ are given as follows:
    \begin{align*}
    A=\begin{pmatrix}
    1.01 & 0 \\
    0 & 1.01
    \end{pmatrix}, \quad \quad C_{i}=\begin{pmatrix} 2\upsilon_{i} & 0 \\ 0 & 2\upsilon_{i} \end{pmatrix},\end{align*}
    \begin{align*} Q=\begin{pmatrix} 0.00002
    & 0 \\ 0 & 0.00002 \end{pmatrix},\quad  R_{i}=\begin{pmatrix}0.5 & 0 \\ 0 & 0.5\end{pmatrix},
    \end{align*}
   $\Gamma=
    \left(
      \begin{array}{ccccc}
        1 & 1 & 0 & 0 & 0 \\
        1 & 1 & 0 & 1 & 0 \\
        1 & 0 & 1 & 0 & 0 \\
        0 & 0 & 1 & 1 & 0 \\
        0 & 1 & 0 & 0 & 1 \\
      \end{array}
    \right)$, where random variable $\upsilon_{i}\in(0,1]$. In this example, we consider the case that measurement noise of each sensor is identical, i.e., $R_{i}=R_{j}$, for all $i,j\in \mathcal{V}$.
    First, we show the asymptotically convergence of the trace of estimation error covariance matrix of each node in Fig.~\ref{1} and Fig.~\ref{2}, which adopts randomized consensus Kalman filtering algorithm in the WSNs.    As depicted in Fig.~\ref{1} and Fig.~\ref{2}, the estimation error, i.e., the trace of error covariance matrix of each node is asymptotically stable when time is long enough. Furthermore, we also compare the performance of different estimation algorithms with different numbers of consensus iteration $K$. Here $K$ is varied from $1$ to $41$ at increments of $5$. Other parameters are kept constant and the priors are chosen to be equal.
    \begin{figure}
    	\centering
    	\includegraphics[width=0.5\textwidth]{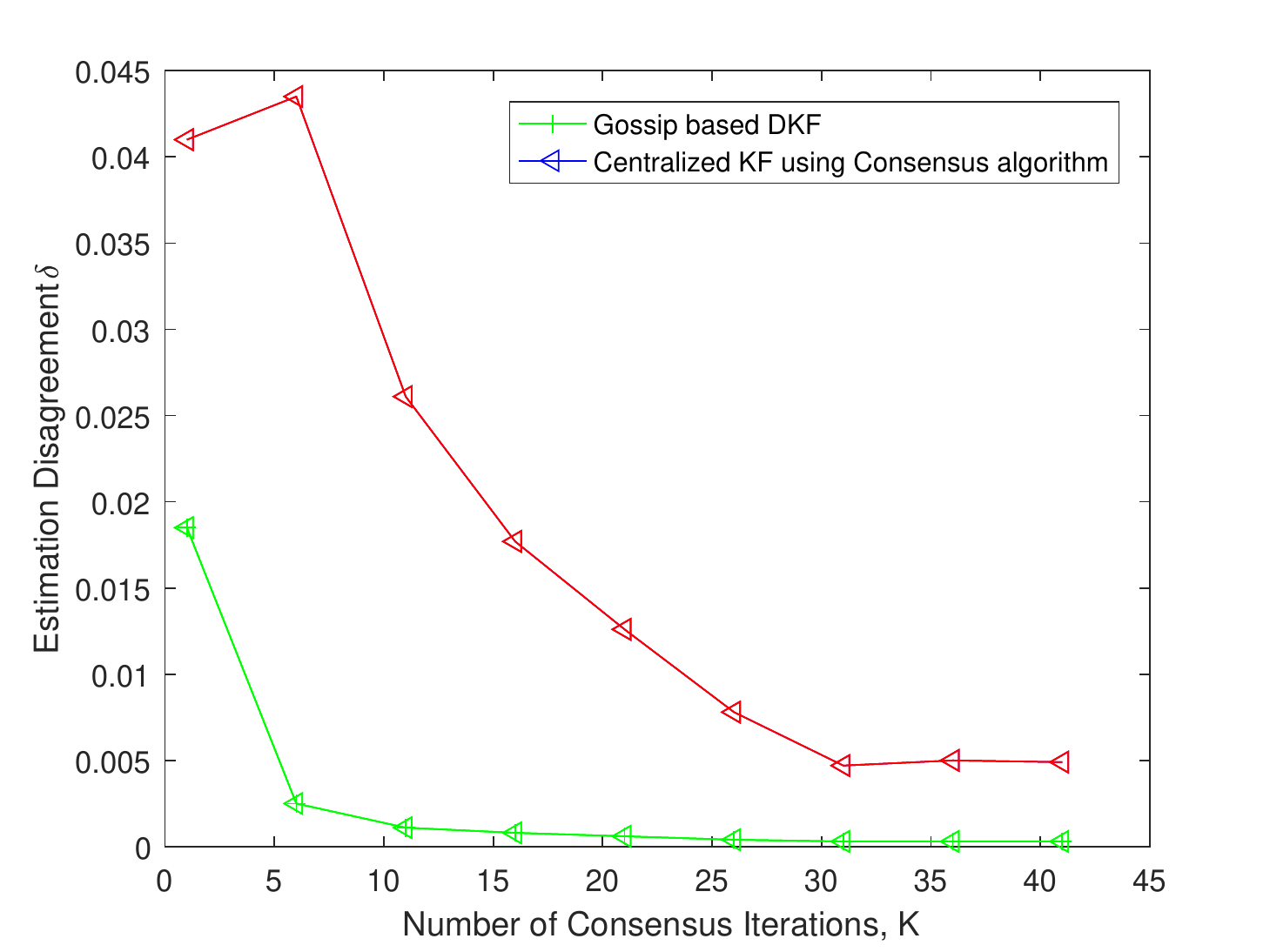}\\
    	\caption{Compare performance of different iteration $K$}\label{iteration}
    \end{figure}
 \begin{figure}
	\centering
	\includegraphics[width=0.5\textwidth]{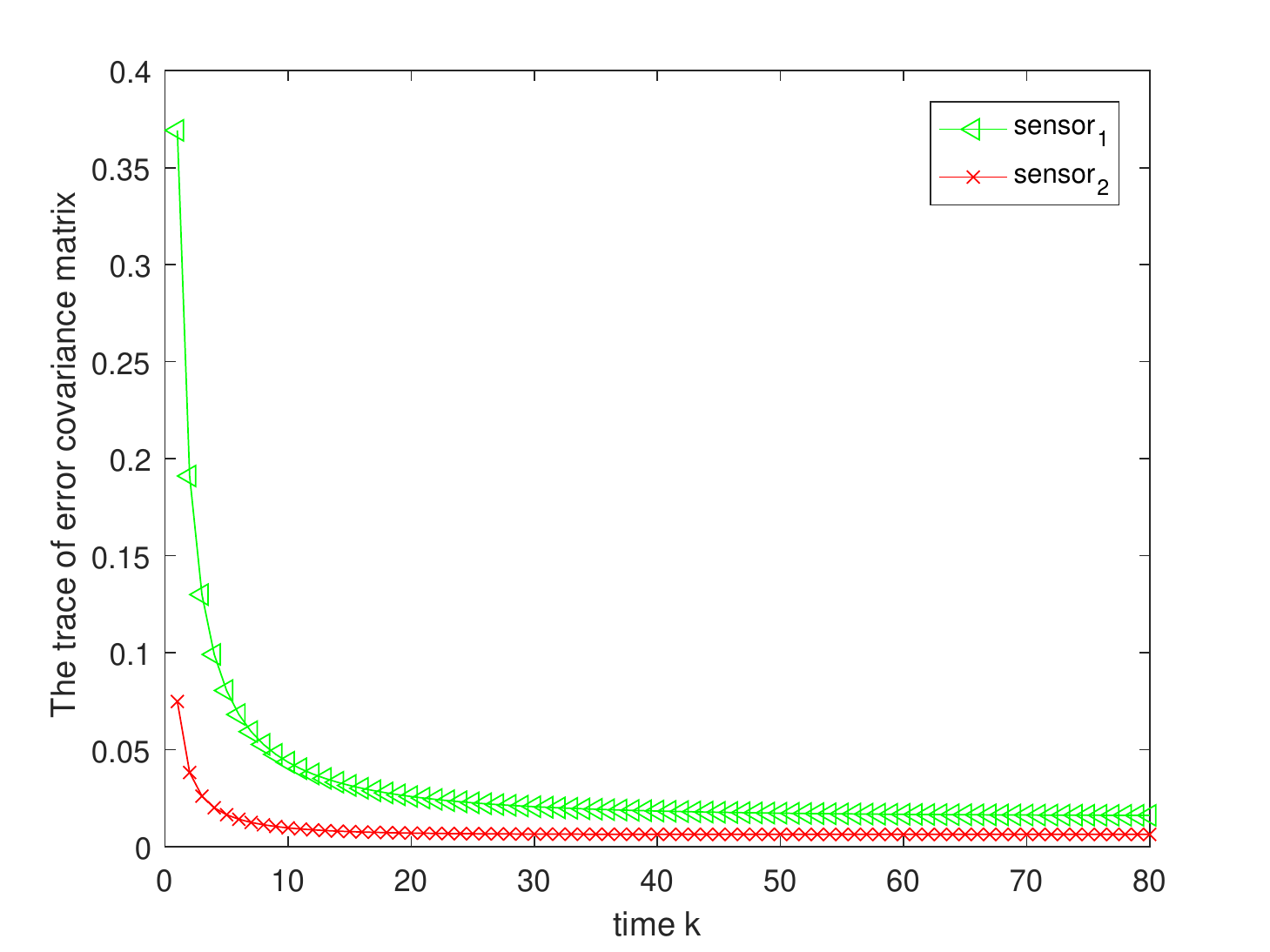}\\
	\caption{The trace of error covariance matrix of sensors: sensor 1 and sensor 2.}\label{1}
\end{figure}
\begin{figure}
	\centering
	\includegraphics[width=0.5\textwidth]{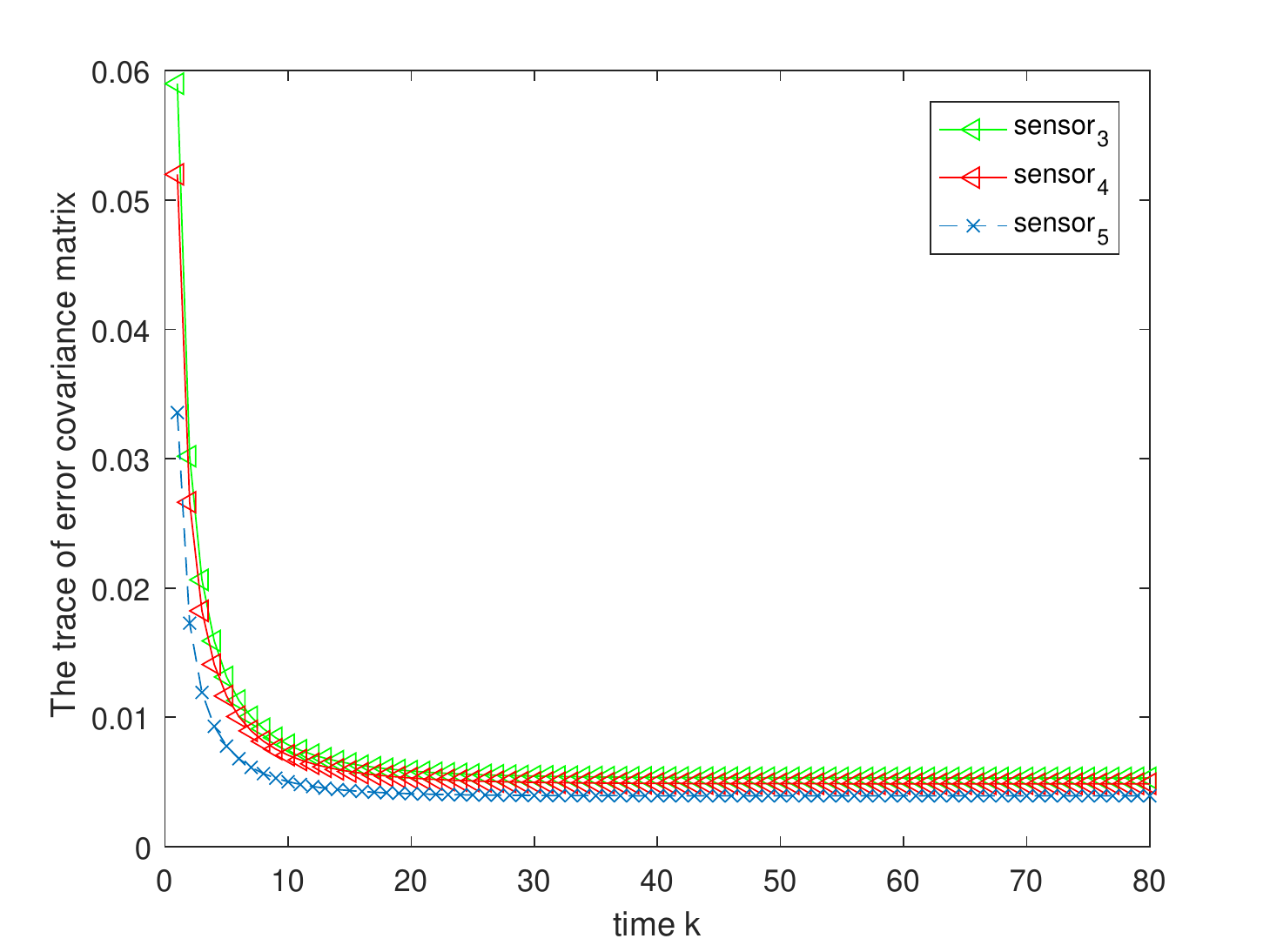}\\
	\caption{The trace of error covariance matrix of sensors: sensor 3, sensor 4, and sensor 5.}\label{2}
\end{figure}
\begin{figure}
       \centering
       \includegraphics[width=0.5\textwidth]{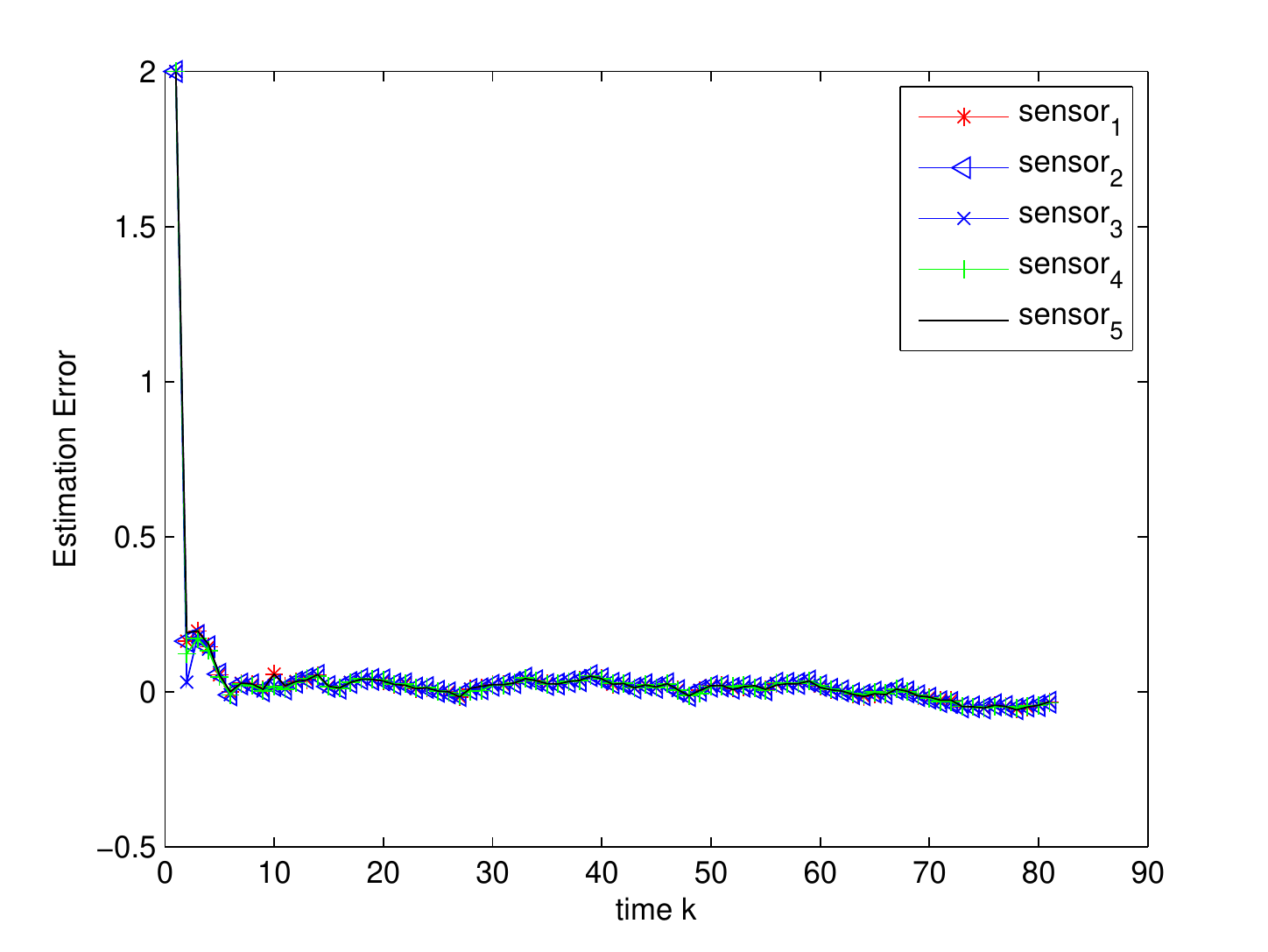}
        \caption{Estimation error per node using randomized gossip based distributed Kalman filtering algorithm. All the curves are obtain after averaged by 100 times.}\label{a}
\end{figure}

	Our next step is to define the mean-square estimation error of tracking state (MSEE) for sensor $i$ at time $k$ as
	\begin{equation}
	MSEE_{i}(k)=||x(k)-\hat{x}^{i}_{k}||^{2}.
	\end{equation}
   In Fig.~\ref{fig:state_error}, we show the mean-square estimation error of tracking state per node adopting randomized gossip based distributed Kalman filtering algorithm.  Fig. \ref{fig:no_consensus} depicts the mean-square estimation error per node without cooperation steps. As depicted in Figs. \ref{fig:state_error} and \ref{fig:no_consensus}, Fig. \ref{fig:state_error} has a coincident state performance due to cooperation steps while there exists disagreement among sensors in Fig.~ \ref{fig:no_consensus}.
   \begin{figure}
   	\centering
   	\includegraphics[width=0.5\textwidth]{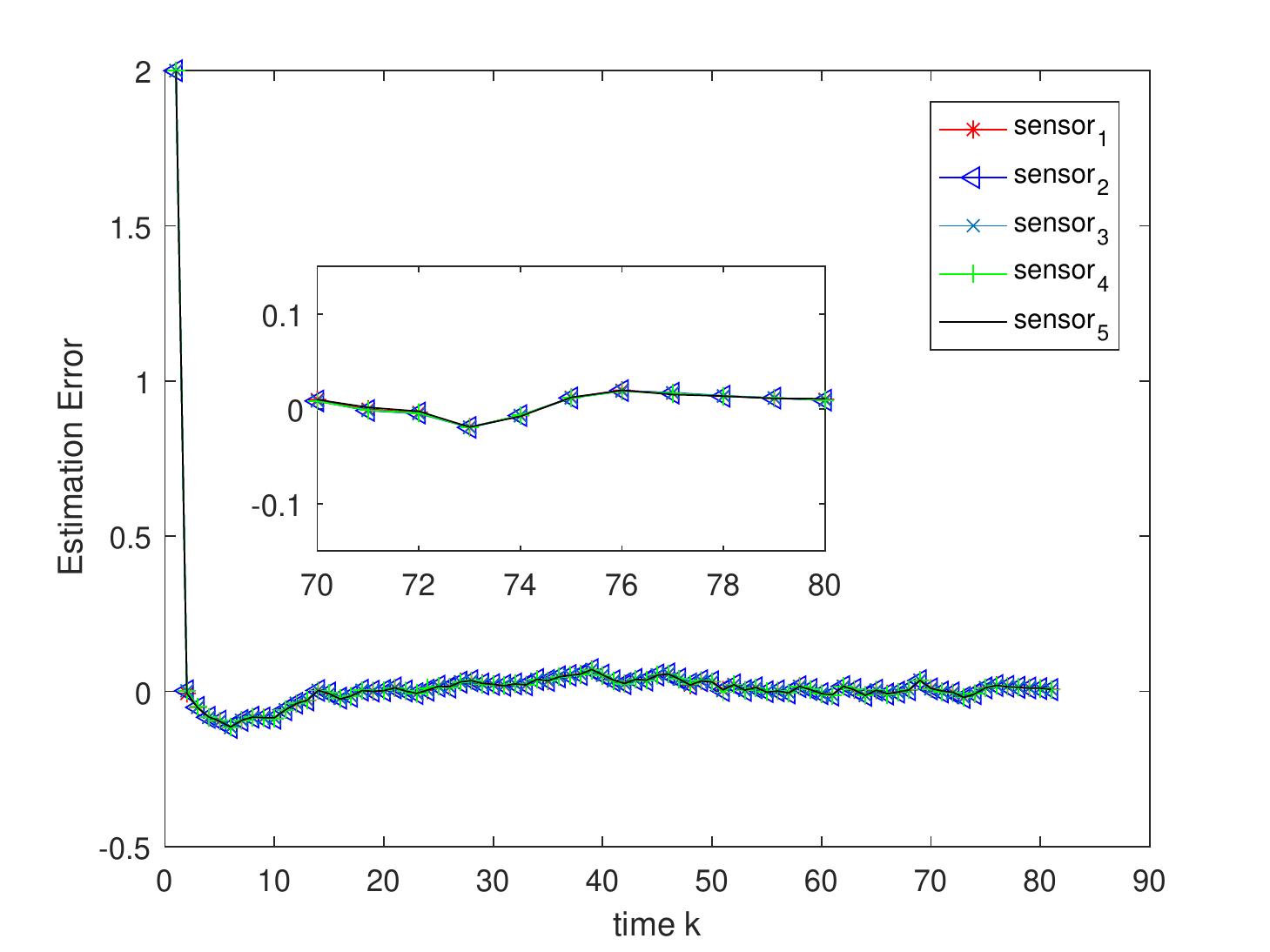}\\
   	\caption{Estimation error of tracking state per node using gossip based distributed Kalman filtering algorithm. All the curves are obtained after being averaged by 100 times.}\label{fig:state_error}
   \end{figure}

	\begin{figure}
	  \centering
	  \includegraphics[width=0.5\textwidth]{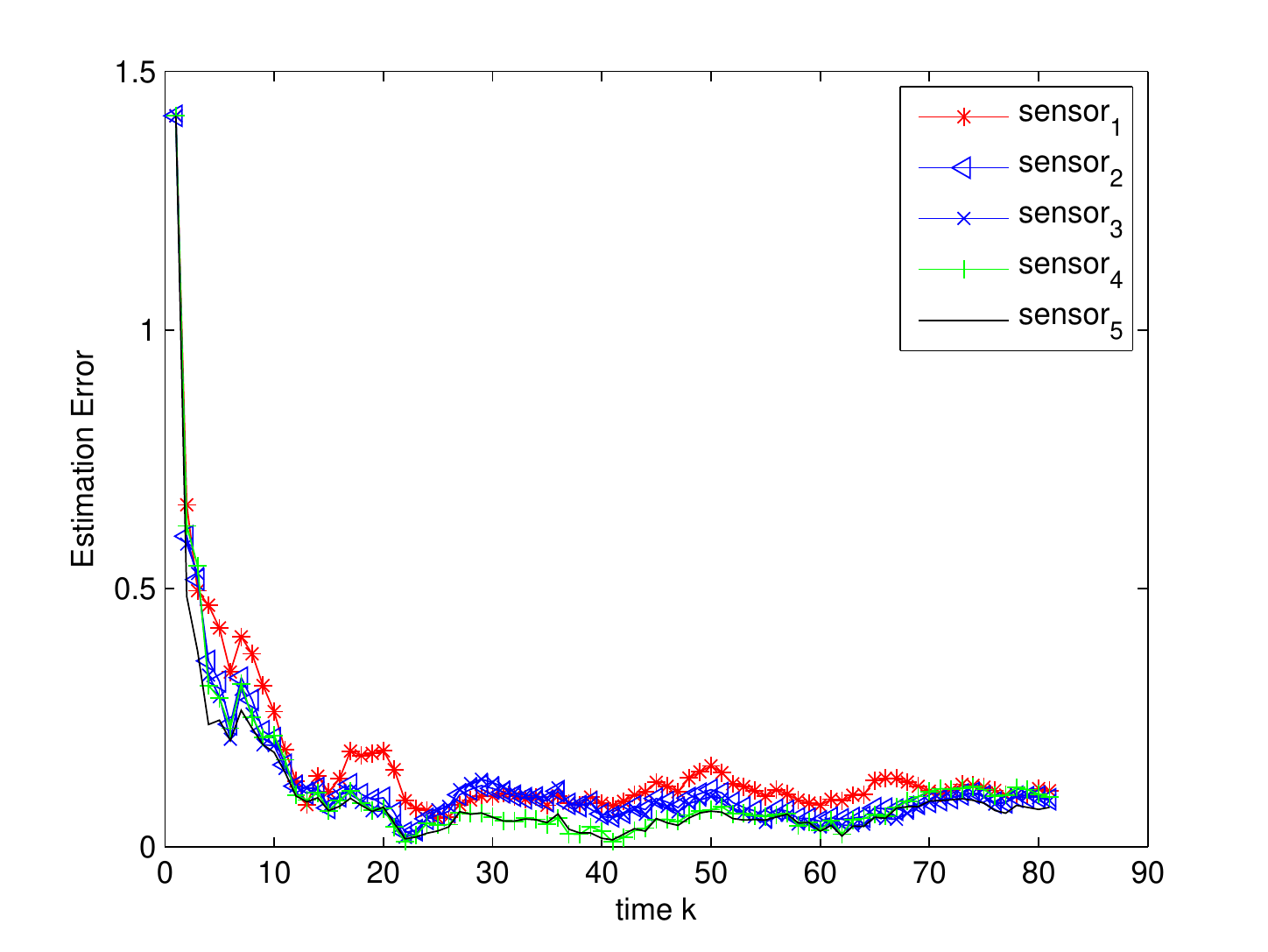}\\
	  \caption{Estimation error per node without cooperation step}\label{fig:no_consensus}
	\end{figure}
    \begin{figure}
    	\centering
    	\includegraphics[width=0.5\textwidth]{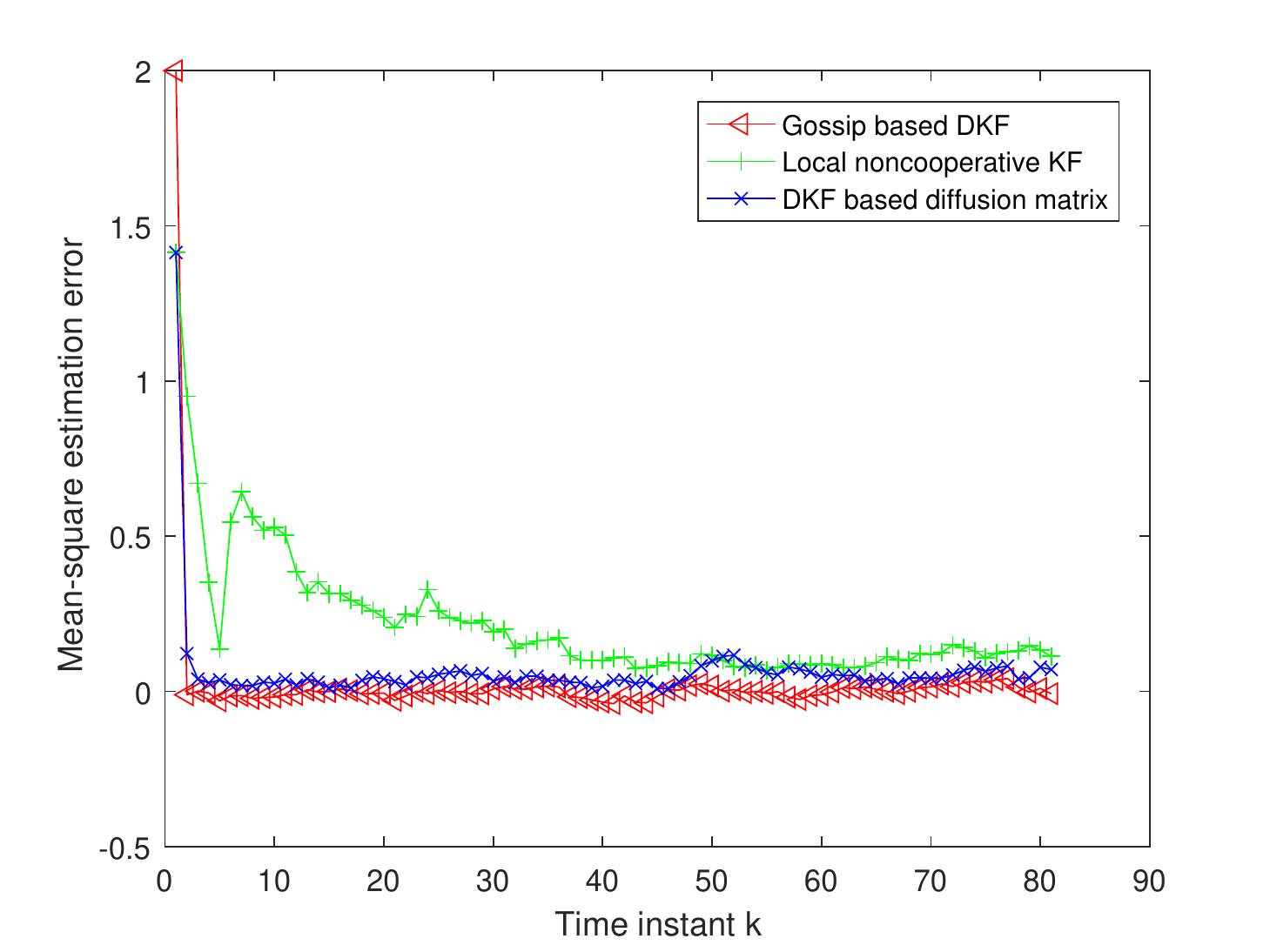}\\
    	\caption{The performance compare of randomized gossip based distributed Kalman filtering algorithm, diffusion strategy based Kalman filtering algorithm, and local noncooperative Kalman filtering algorithm. All the curves are obtained after being averaged by 100 times.}\label{d}
    \end{figure}
    Fig. \ref{d} compares the performance of different cases including randomized gossip consensus, diffusion strategy, and noncooperation strategy based distributed Kalman filtering algorithm, respectively.
    Moreover, for the comparison of three distributed Kalman filtering methods in detail, the average
    mean-square estimation error (MSEE) of $n$ sensors tracking state error are further employed to analyze the performance and defined as follows:
    \begin{equation}
    MSEE_{ave}(k)=\frac{1}{n}\sum_{i=1}^{n}MSEE_{i}(k).
    \end{equation}
    As depicted in Fig. \ref{d}, all the results are obtained after being averaged by 100 times. The proposed randomized gossip based Kalman filtering algorithm has smaller average mean-square estimation error than other two methods.

	Finally, we also show the average disagreement of state tracking by adopting our proposed algorithm, diffusion strategy based Kalman filtering algorithm, and local noncooperative Kalman filtering algorithm in Fig.~\ref{c}.
    To measure the disagreement of the estimates independent of network topology, we evaluate the performance of the algorithm
    by a normalized version of the distance from the consensus value as follows:
    \begin{equation}
    ||\delta||=(\sum_{i=1}^{n}||\hat{x}_{i}-x_{A}||_{2}^{2})^{1/2}
    \end{equation}
    with $x_{A}=\frac{1}{n}\sum_{i=1}^{n}\hat{x}_{i}$.
     \begin{figure}[ht]
    	\centering
    	\includegraphics[width=0.5\textwidth]{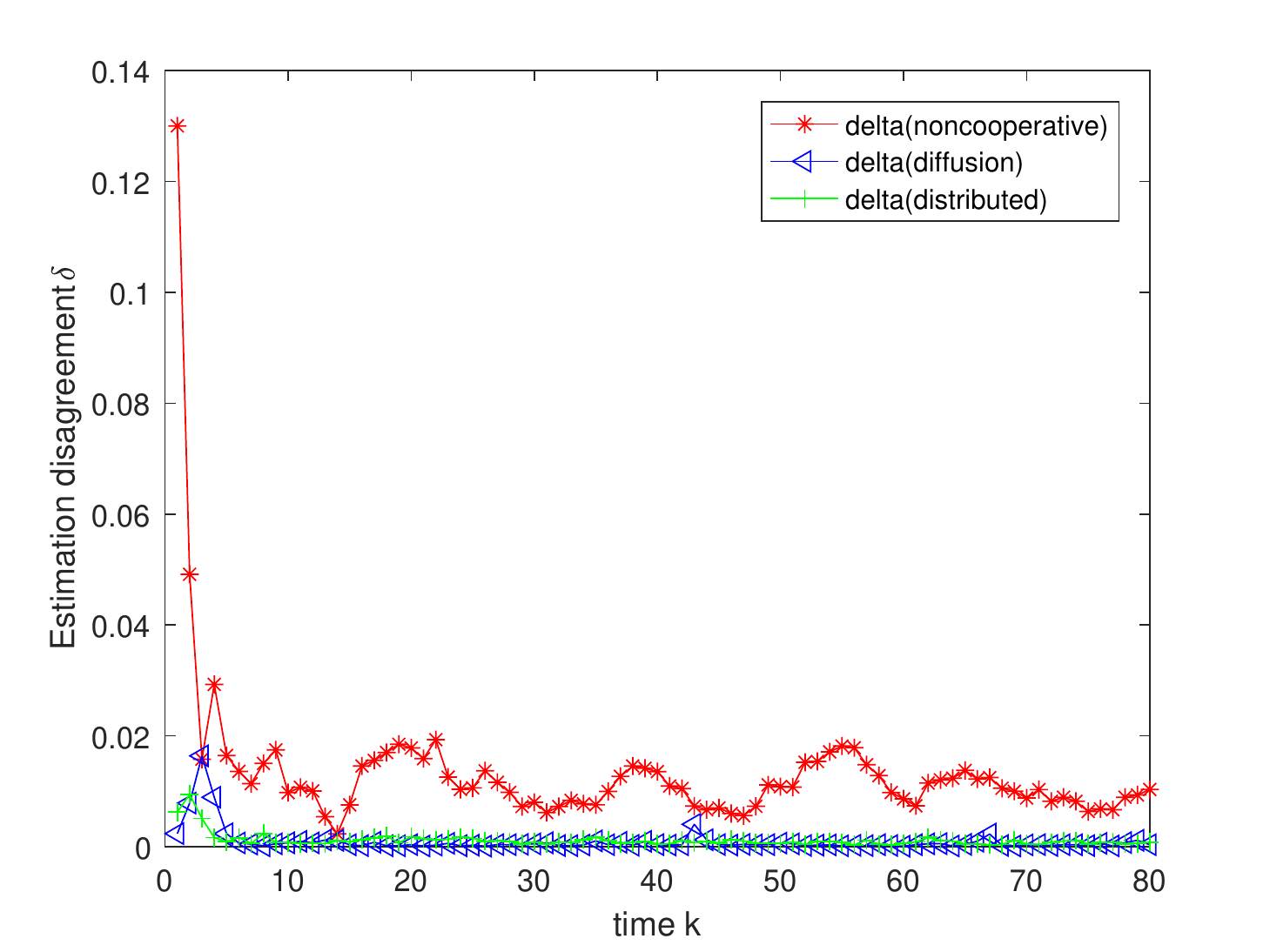}\\
    	\caption{Disagreement Estimates.  All the curves are obtain after averaged by 100 times.}\label{c}
    \end{figure}
	As depicted in Fig. \ref{c}, the estimation disagreement of gossip strategy and diffusion strategy based Kalman filtering algorithm are smaller than local noncooperative Kalman filtering algorithm due to consensus steps.

\end{example}

\section{Conclusions}\label{sec:conclusion}
In this paper, we have proposed a novel algorithm called randomized gossip based distributed Kalman filtering algorithm. Under some mild assumption, we have obtained a sufficient condition for guaranteeing the convergence of the expected estimation error covariance for the proposed algorithm. We also investigate the sensor scheduling problem for distributed estimation when adopting randomized gossip based distributed Kalman filtering algorithm under the power constraint. By relaxing the optimal sensor scheduling problem to a convex optimization problem in a set of linear matrix inequalities, we have provided the sub-optimal solution. The simulation results verify that the better estimates can be obtained when compared with the diffusion based distributed Kalman filtering algorithm and noncooperative decentralized Kalman filtering algorithm. Besides, the average disagreement of estimates of our algorithm is smaller than noncooperative decentralized Kalman filtering algorithm and diffusion strategy based Kalman filtering algorithm.
In our current work, we investigate the scheduling problem when adopting proposed algorithm in a reliable channel.  However, in practical applications, channels may be unreliable and have packet delays or droppings. As a future work, we will study  other general communication channel models including packet-delay or fading ones.

\begin{appendices}
	\section{Proof of Theorem 4.1}
	       The iteration is obvious. The estimation error covariance $\mathcal{P}_{\widetilde{x},k}$ depends upon the $\mathcal{W}$ which is randomly selected from the set at every time $k$. Thus, we
	show that the estimation error covariance of our proposed algorithm converge in expectation to the stable state as follows.
	
	We focus on the proof of the fixed point of the error covariance matrix. For any $X,Y \in R^{nm \times nm}$, we have
	\begin{equation*}
	||T(X)-T(Y)||^{2}=Tr((T(X)-T(Y))(T(X)-T(Y))'),
	\end{equation*}
	where $||\cdot||$ represents Frobenius norm. Denote
	$\mathcal{\bar{W}}=E[\mathcal{W}]=E[W(k)]\otimes I_{m}=W \otimes I_{m}$, where $W$ is given in equation \eqref{23}.
	From the discussion above, we know that
	\begin{equation*}
	\begin{aligned}
	&||T(X)-T(Y)||^{2}=Tr((\mathcal{\bar{W}}[\mathcal{\bar{A}}X \mathcal{\bar{A}}'+\mathcal{\bar{B}}(\mathbf{1}\mathbf{1}'\otimes Q)\mathcal{\bar{B}}'+\\
	&\mathcal{\bar{D}}R\mathcal{\bar{D}}']\mathcal{\bar{W}}'-\mathcal{\bar{W}}[\mathcal{\bar{A}}Y \mathcal{\bar{A}}'+\mathcal{\bar{B}}(\mathbf{1}\mathbf{1}'\otimes Q)\mathcal{\bar{B}}'+\mathcal{\bar{D}}R\mathcal{\bar{D}}']\mathcal{\bar{W}}'
	)(\cdot)').
	\end{aligned}
	\end{equation*}
	Define $ \varGamma(X)=\mathcal{\bar{A}}X \mathcal{\bar{A}}'+\mathcal{\bar{B}}(\mathbf{1}\mathbf{1}'\otimes Q)\mathcal{\bar{B}}'+\mathcal{\bar{D}}R\mathcal{\bar{D}}' $ and
	$\varGamma(Y)=\mathcal{\bar{A}}Y \mathcal{\bar{A}}'+\mathcal{\bar{B}}(\mathbf{1}\mathbf{1}'\otimes Q)\mathcal{\bar{B}}'+\mathcal{\bar{D}}R\mathcal{\bar{D}}'$.
	Then, we have
	\begin{equation*}
	\begin{aligned}
	&Tr((T(X)-T(Y))(T(X)-T(Y))')\\
	&=Tr((\mathcal{\bar{W}}(\varGamma(X)
	-\varGamma(Y))\mathcal{\bar{W}}')(\mathcal{\bar{W}}(\varGamma(X)-\varGamma(Y))\mathcal{\bar{W}}')')\\
	&\overset{a}{=}Tr(\mathcal{\bar{W}}\mathcal{\bar{W}}'((\varGamma(X)-\varGamma(Y))(\varGamma(X)-\varGamma(Y))^{'}\mathcal{\bar{W}}^{'}\mathcal{\bar{W}}).
	\end{aligned}
	\end{equation*}
	The equality (a) is derived from the fact that  $Tr(ABC)=Tr(ACB)=Tr(BAC)=Tr(BCA)=Tr(CAB)=Tr(CBA).$
	Note that the estimation error covariance matrix of the decentralized Kalman filtering converges to a steady state.
	More specifically, when $W$ is always an identity matrix, $\varGamma$ is a contraction mapping due to the convergence of decentralized Kalman filtering algorithm.
	\begin{equation*}
	Tr((\Gamma(Y)-\Gamma(X)(\Gamma(Y)-\Gamma(X)')
	\leq  \rho ||Y-X||,
	\end{equation*}
	where $\rho\in(0,1)$.
	Thus, according to the above discussions, we have that
	\begin{equation}\label{52}
	\begin{aligned}
	&||T(Y)-T(X)||^{2}=Tr((T(X)-T(Y))(T(X)-T(Y))')\\
	&=Tr(\mathcal{\bar{W}}\mathcal{\bar{W}}'((\varGamma(X)-\varGamma(Y))(\varGamma(X)-\varGamma(Y))^{'}\mathcal{\bar{W}}^{'}\mathcal{\bar{W}})\\
	&\overset{b}{\leq} Tr((\Gamma(Y)-\Gamma(X)(\Gamma(Y)-\Gamma(X)')
	\leq  \rho ||Y-X||.
	\end{aligned}
	\end{equation}
	The inequality (b) is derived from the lemma \ref{lemma1} based on the fact that $(\Gamma(Y)-\Gamma(X))(\Gamma(Y)-\Gamma(X))'$ is a symmetric positive-definite matrix and $\mathcal{\bar{W}}$ is a symmetric stochastic matrix.  The inequality \eqref{52} illustrates that $T$ is a contraction mapping and a unique fixed point exists for the mapping $T$ according to the Banach fixed point theorem. Thus the conclusion follows.
	\section{Proof of Theorem 4.2}
	 We demonstrate that $Tr(E[\mathcal{P}_{\widetilde{x},k}]) \leqslant Tr(\mathcal{P}_{k})$,  $\forall k\in R$ by induction as follows. First note that the initial condition $\mathcal{P}_{0}=\mathcal{P}_{\widetilde{x},0} \geq 0$. Thus $Tr(E(\mathcal{P}_{0}))=Tr(\mathcal{\hat{P}}_{0})$ is tenable at time $k=0$.
	Then assume that at time $k$, there holds $Tr(E[\mathcal{P}_{\widetilde{x},k}]) \leqslant Tr(\mathcal{P}_{k})$.
	What we need to prove is that $Tr(E[\mathcal{P}_{\widetilde{x},k+1}])\leqslant Tr(\mathcal{P}_{k+1})$ also holds at time $k+1$.
	
	We have that $E(\mathcal{P}_{\widetilde{x},k+1})=E(\mathcal{W}\mathcal{\bar{A}}\mathcal{P}_{\widetilde{x},k}\mathcal{\bar{A}}'\mathcal{W}'+\mathcal{W}\mathcal{\bar{B}}(\mathbf{1}\mathbf{1}'\otimes Q)\mathcal{\bar{B}}'\mathcal{W}'+\mathcal{W}\mathcal{\bar{D}}R\mathcal{\bar{D}}'\mathcal{W}')$.
	By using the full probability formula based on the conditional expectation, we obtain that $E(\mathcal{P}_{\widetilde{x},k+1})=E(E(\mathcal{P}_{\widetilde{x},k+1}|\mathcal{P}_{\widetilde{x},k}))$, and $
	E(E(\mathcal{P}_{\widetilde{x},k+1}|\mathcal{P}_{\widetilde{x},k}))=E[\mathcal{W}\mathcal{\bar{A}}E(\mathcal{P}_{\widetilde{x},k})\mathcal{\bar{A}}'\mathcal{W}'+
	\mathcal{W}\mathcal{\bar{B}}(\mathbf{1}\mathbf{1}'\otimes Q)\mathcal{\bar{B}}'\mathcal{W}'
	+\mathcal{W}\mathcal{\bar{D}}R\mathcal{\bar{D}}'\mathcal{W}']$. Combining these relations, we have
	\begin{equation*}
	\begin{aligned}
	&Tr(E(\mathcal{P}_{\widetilde{x},k+1}))=Tr(E[E(\mathcal{W}\mathcal{\bar{A}}\mathcal{P}_{\widetilde{x},k}\mathcal{\bar{A}}'\mathcal{W}'
	+\mathcal{W}\mathcal{\bar{B}}(\mathbf{1}\mathbf{1}' \\
	&\otimes Q) \mathcal{\bar{B}}'\mathcal{W}'+\mathcal{W}\mathcal{\bar{D}}R\mathcal{\bar{D}}'\mathcal{W}')])=Tr(\mathcal{W}\mathcal{\bar{A}}E(\mathcal{P}_{\widetilde{x},k})\mathcal{\bar{A}}'\mathcal{W}'\\
	&+\mathcal{W}\mathcal{\bar{B}}(\mathbf{1}\mathbf{1}'\otimes Q)\mathcal{\bar{B}}'\mathcal{W}'+\mathcal{W}\mathcal{\bar{D}}R\mathcal{\bar{D}}'\mathcal{W}').
	\end{aligned}\end{equation*}
	Now,  $\mathcal{\bar{A}}E(\mathcal{P}_{\widetilde{x},k})\mathcal{\bar{A}}'+
	\mathcal{\bar{B}}(\mathbf{1}\mathbf{1}'\otimes Q)\mathcal{\bar{B}}'
	+\mathcal{\bar{D}}R\mathcal{\bar{D}}'$ is a positive-definite matrix due to the fact that $\mathcal{P}_{\widetilde{x},k},Q,R$ are all positive-definite matrices, respectively.
	Based on the \textit{Lemma \ref{lemma1}}, we further have that
	\begin{align*}
	&Tr(\mathcal{W}\mathcal{\bar{A}}E(\mathcal{P}_{\widetilde{x},k})\mathcal{\bar{A}}'\mathcal{W}'+
	\mathcal{W}\mathcal{\bar{B}}(\mathbf{1}\mathbf{1}'\otimes Q)\mathcal{\bar{B}}'\mathcal{W}'
	+\mathcal{W}\mathcal{\bar{D}}R\\
	&\mathcal{\bar{D}}'\mathcal{W}') \leqslant Tr(\mathcal{\bar{A}}E(\mathcal{P}_{\widetilde{x},k})\mathcal{\bar{A}}'+
	\mathcal{\bar{B}}(\mathbf{1}\mathbf{1}'\otimes Q)\mathcal{\bar{B}}'
	+\mathcal{\bar{D}}R\mathcal{\bar{D}}').
	\end{align*}
	In view of the assumption that $Tr(E[\mathcal{P}_{\widetilde{x},k}) \leqslant Tr(\mathcal{P}_{k})$ and  $\mathcal{\bar{A}}=\mathcal{P}(\mathcal{P}^{-})^{-1}(I \otimes A)$ is an orthogonal matrix. There holds that $\mathcal{\bar{A}}E(\mathcal{P}_{\widetilde{x},k})\mathcal{\bar{A}}'=\mathcal{\bar{A}}E(\mathcal{P}_{\widetilde{x},k})\mathcal{\bar{A}}^{-1}$ and $Tr(\mathcal{\bar{A}}E(\mathcal{P}_{\widetilde{x},k})\mathcal{\bar{A}}')=Tr(E(\mathcal{P}_{\widetilde{x},k}))\leqslant Tr(\mathcal{P}_{k})=Tr(\mathcal{\bar{A}}\mathcal{{P}}_{k}\mathcal{\bar{A}}')$.
	Note that
	\begin{align*}
	&Tr(\mathcal{\bar{A}}E(\mathcal{P}_{\widetilde{x},k})\mathcal{\bar{A}}'+\mathcal{\bar{B}}(\mathbf{1}\mathbf{1}'\otimes Q)\mathcal{\bar{B}}'
	+\mathcal{\bar{D}}R\mathcal{\bar{D}}')
	\leqslant E[Tr(\mathcal{\bar{A}}\\
	&\mathcal{{P}}_{k}\mathcal{\bar{A}}'+
	\mathcal{\bar{B}}(\mathbf{1}\mathbf{1}'\otimes Q)\mathcal{\bar{B}}'+\mathcal{\bar{D}}R\mathcal{\bar{D}}')]\\
	&= E[Tr(\mathcal{{P}}_{k+1})]\\
	&=Tr(\mathcal{P}_{k+1}).
	\end{align*}
	By inductive hypothesis, $Tr(E(\mathcal{P}_{\widetilde{x},k}))\leqslant Tr(\mathcal{P}_{k})$ holds for all time instant $k\geq0$.
	The proof is thus completed.
\end{appendices}

\bibliographystyle{IEEE}

\end{document}